\documentclass[twocolumn]{aastex631}
\usepackage{amsthm,amsmath,amssymb}
\usepackage{graphicx}
\usepackage{xcolor}
\usepackage{float} 
\usepackage{comment}
\usepackage{natbib}

\newcommand{\revision}{}

\newcommand{\e}{\text{e}}

\newcommand{\p}{\partial}

\newcommand{\yisx}{\color{black}}
\newcommand{\yisxapj}{\color{black}}

\begin{document}

\title{Evidence of mini-jet emission in a large emission zone from a magnetically-dominated gamma-ray burst jet}

\author{S.-X. Yi}
\affil{Key Laboratory of Particle Astrophysics, Institute of High Energy Physics, Chinese Academy of Sciences, Beijing 100049, China}
\email{sxyi@ihep.ac.cn}

\author{C.-W. Wang}
\affil{Key Laboratory of Particle Astrophysics, Institute of High Energy Physics, Chinese Academy of Sciences, Beijing 100049, China}
\affil{University of Chinese Academy of Sciences, Chinese Academy of Sciences, Beijing 100049, China}
\email{cwwang@ihep.ac.cn}

\author{X. Shao}
\affil{Department of Astronomy, Beijing Normal University, Beijing 100875, People’s Republic of China}

\author{R. Moradi}
\affil{Key Laboratory of Particle Astrophysics, Institute of High Energy Physics, Chinese Academy of Sciences, Beijing 100049, China}

\author{H. Gao}
\affil{Department of Astronomy, Beijing Normal University, Beijing 100875, People’s Republic of China}
\email{gaohe@bnu.edu.cn}

\author{B. Zhang}
\affil{Nevada Center for Astrophysics, University of Nevada Las Vegas, NV 89154, USA}
\affil{Department of Physics and Astronomy, University of Nevada Las Vegas, NV 89154, USA}

\author{S.-L. Xiong}
\affil{Key Laboratory of Particle Astrophysics, Institute of High Energy Physics, Chinese Academy of Sciences, Beijing 100049, China}
\email{xiongsl@ihep.ac.cn}

\author{S.-N. Zhang}
\affil{Key Laboratory of Particle Astrophysics, Institute of High Energy Physics, Chinese Academy of Sciences, Beijing 100049, China}
\affil{University of Chinese Academy of Sciences, Chinese Academy of Sciences, Beijing 100049, China}

\author{W.-J. Tan}
\affil{Key Laboratory of Particle Astrophysics, Institute of High Energy Physics, Chinese Academy of Sciences, Beijing 100049, China}
\affil{University of Chinese Academy of Sciences, Chinese Academy of Sciences, Beijing 100049, China}

\author{J.-C. Liu}
\affil{Key Laboratory of Particle Astrophysics, Institute of High Energy Physics, Chinese Academy of Sciences, Beijing 100049, China}
\affil{University of Chinese Academy of Sciences, Chinese Academy of Sciences, Beijing 100049, China}

\author{W.-C. Xue}
\affil{Key Laboratory of Particle Astrophysics, Institute of High Energy Physics, Chinese Academy of Sciences, Beijing 100049, China}
\affil{University of Chinese Academy of Sciences, Chinese Academy of Sciences, Beijing 100049, China}

\author{Y.-Q. Zhang}
\affil{Key Laboratory of Particle Astrophysics, Institute of High Energy Physics, Chinese Academy of Sciences, Beijing 100049, China}
\affil{University of Chinese Academy of Sciences, Chinese Academy of Sciences, Beijing 100049, China}

\author{C. Zheng}
\affil{Key Laboratory of Particle Astrophysics, Institute of High Energy Physics, Chinese Academy of Sciences, Beijing 100049, China}
\affil{University of Chinese Academy of Sciences, Chinese Academy of Sciences, Beijing 100049, China}

\author{Y. Wang}
\affil{ICRANet, Piazza della Repubblica 10, I-65122 Pescara, Italy}
\affil{ICRA, Dipartamento di Fisica, Sapienza Universit\`a  di Roma, Piazzale Aldo Moro 5, I-00185 Rome, Italy}
\affil{INAF, Osservatorio Astronomico d'Abruzzo, Via M. Maggini snc, I-64100, Teramo, Italy}

\author{P. Zhang}
\affil{Key Laboratory of Particle Astrophysics, Institute of High Energy Physics, Chinese Academy of Sciences, Beijing 100049, China}
\affil{College of Electronic and Information Engineering, Tongji University, Shanghai 201804, China}

\author{Z.-H. An}
\affil{Key Laboratory of Particle Astrophysics, Institute of High Energy Physics, Chinese Academy of Sciences, Beijing 100049, China}

\author{C. Cai}
\affil{College of Physics and Hebei Key Laboratory of Photophysics Research and Application, Hebei Normal University, Shijiazhuang, Hebei 050024, China}

\author{P.-Y. Feng}
\affil{Key Laboratory of Particle Astrophysics, Institute of High Energy Physics, Chinese Academy of Sciences, Beijing 100049, China}
\affil{University of Chinese Academy of Sciences, Chinese Academy of Sciences, Beijing 100049, China}

\author{K. Gong}
\affil{Key Laboratory of Particle Astrophysics, Institute of High Energy Physics, Chinese Academy of Sciences, Beijing 100049, China}

\author{D.-Y. Guo}
\affil{Key Laboratory of Particle Astrophysics, Institute of High Energy Physics, Chinese Academy of Sciences, Beijing 100049, China}

\author{Y. Huang}
\affil{Key Laboratory of Particle Astrophysics, Institute of High Energy Physics, Chinese Academy of Sciences, Beijing 100049, China}

\author{B. Li}
\affil{Key Laboratory of Particle Astrophysics, Institute of High Energy Physics, Chinese Academy of Sciences, Beijing 100049, China}

\author{X.-B. Li}
\affil{Key Laboratory of Particle Astrophysics, Institute of High Energy Physics, Chinese Academy of Sciences, Beijing 100049, China}

\author{X.-Q. Li}
\affil{Key Laboratory of Particle Astrophysics, Institute of High Energy Physics, Chinese Academy of Sciences, Beijing 100049, China}

\author{X.-J. Liu}
\affil{Key Laboratory of Particle Astrophysics, Institute of High Energy Physics, Chinese Academy of Sciences, Beijing 100049, China}

\author{Y.-Q. Liu}
\affil{Key Laboratory of Particle Astrophysics, Institute of High Energy Physics, Chinese Academy of Sciences, Beijing 100049, China}

\author{X. Ma}
\affil{Key Laboratory of Particle Astrophysics, Institute of High Energy Physics, Chinese Academy of Sciences, Beijing 100049, China}

\author{W.-X. Peng}
\affil{Key Laboratory of Particle Astrophysics, Institute of High Energy Physics, Chinese Academy of Sciences, Beijing 100049, China}

\author{R. Qiao}
\affil{Key Laboratory of Particle Astrophysics, Institute of High Energy Physics, Chinese Academy of Sciences, Beijing 100049, China}

\author{L.-M. Song}
\affil{Key Laboratory of Particle Astrophysics, Institute of High Energy Physics, Chinese Academy of Sciences, Beijing 100049, China}

\author{J. Wang}
\affil{Key Laboratory of Particle Astrophysics, Institute of High Energy Physics, Chinese Academy of Sciences, Beijing 100049, China}

\author{P. Wang}
\affil{Key Laboratory of Particle Astrophysics, Institute of High Energy Physics, Chinese Academy of Sciences, Beijing 100049, China}

\author{Y. Wang}
\affil{Key Laboratory of Particle Astrophysics, Institute of High Energy Physics, Chinese Academy of Sciences, Beijing 100049, China}
\affil{University of Chinese Academy of Sciences, Chinese Academy of Sciences, Beijing 100049, China}

\author{X.-Y. Wen}
\affil{Key Laboratory of Particle Astrophysics, Institute of High Energy Physics, Chinese Academy of Sciences, Beijing 100049, China}

\author{S. Xiao}
\affil{Guizhou Provincial Key Laboratory of Radio Astronomy and Data Processing, Guizhou Normal University, Guiyang 550001, China}

\author{Y.-B. Xu}
\affil{Key Laboratory of Particle Astrophysics, Institute of High Energy Physics, Chinese Academy of Sciences, Beijing 100049, China}

\author{S. Yang}
\affil{Key Laboratory of Particle Astrophysics, Institute of High Energy Physics, Chinese Academy of Sciences, Beijing 100049, China}

\author{Q.-B. Yi}
\affil{Key Laboratory of Particle Astrophysics, Institute of High Energy Physics, Chinese Academy of Sciences, Beijing 100049, China}
\affil{School of Physics and Optoelectronics, Xiangtan University, Yuhu District, Xiangtan, Hunan, 411105, China}

\author{D.-L. Zhang}
\affil{Key Laboratory of Particle Astrophysics, Institute of High Energy Physics, Chinese Academy of Sciences, Beijing 100049, China}

\author{F. Zhang}
\affil{Key Laboratory of Particle Astrophysics, Institute of High Energy Physics, Chinese Academy of Sciences, Beijing 100049, China}

\author{H.-M. Zhang}
\affil{Key Laboratory of Particle Astrophysics, Institute of High Energy Physics, Chinese Academy of Sciences, Beijing 100049, China}

\author{J.-P. Zhang}
\affil{Key Laboratory of Particle Astrophysics, Institute of High Energy Physics, Chinese Academy of Sciences, Beijing 100049, China}
\affil{University of Chinese Academy of Sciences, Chinese Academy of Sciences, Beijing 100049, China}

\author{Z. Zhang}
\affil{Key Laboratory of Particle Astrophysics, Institute of High Energy Physics, Chinese Academy of Sciences, Beijing 100049, China}

\author{X.-Y. Zhao}
\affil{Key Laboratory of Particle Astrophysics, Institute of High Energy Physics, Chinese Academy of Sciences, Beijing 100049, China}

\author{Y. Zhao}
\affil{Key Laboratory of Particle Astrophysics, Institute of High Energy Physics, Chinese Academy of Sciences, Beijing 100049, China}
\affil{School of Computer and Information, Dezhou University, Dezhou 253023, China}

\author{S.-J. Zheng}
\affil{Key Laboratory of Particle Astrophysics, Institute of High Energy Physics, Chinese Academy of Sciences, Beijing 100049, China}
\begin{abstract}
The second brightest GRB in history, GRB230307A, provides an ideal laboratory to study the {\yisx mechanism} of GRB prompt emission thanks to its extraordinarily high photon statistics and its {\yisx single episode activity}. Here we demonstrate that {\yisx the rapidly variable components of its prompt emission compose an overall broad single pulse-like profile. Although these individual rapid components are aligned in time across all energy bands, this overall profile conspires to show a well-defined energy-dependent behavior which is typically seen in single GRB pulses.} Such a feature {\yisx demonstrates that the prompt emission of this burst is from many individual emitting units that are casually linked in a emission site at a large distance from the central engine.} {\yisx Such a scenario is in natural consistency with the internal-collision-induced magnetic reconnection and turbulence framework, which invokes many mini-jets due to local magnetic reconnection that constantly appear and disappear in a global magnetically-dominated jet.}

\end{abstract}

\keywords{Gamma-ray Bursts: Individual: GRB 230307A}

\section{introduction}
The origin of the prompt emission of gamma-ray bursts (GRBs) is still subject to debate because of the not-well-constrained jet composition, location of the emission region, and  mechanism with which $\gamma$-rays are produced \citep{Zhang18}. Depending on the unknown jet composition, three emission sites are commonly discussed in GRB prompt emission models. For a matter-dominated {\yisx outflow}, the observed emission is likely a superposition of a thermal component originating from the fireball photosphere at $R_{\rm{ph}} \sim 10^{11}-10^{12}$ cm \citep{paczynski86,goodman86,meszarosrees00} and a non-thermal component originating from synchrotron-radiating electrons accelerated from internal shocks at $R_{\rm IS} \sim 10^{13}-10^{14}$ cm \citep{Rees-Meszaros94,Kobayashi97,Daigne-Mochkovitch98}. 
There are also models with sub-photospheric dissipiation (see \citealt{2005ApJ...628..847R}). 

{\yisx Alternatively, a GRB jet can be Poynting-flux-dominated which may dissipate its magnetic energy to power radiation. If dissipation occurs at small radii below the photosphere, the magnetic energy dissipation would mostly modify the spectrum of photospheric emission \citep{rees05,thompson07,giannios06,drenkhahn02}. However, if the Poynting-flux energy is retained and gets dissipated only at a large-enough radius well above the photosphere, mini-jets can be generated through magnetic reconnection in a high-$\sigma$ ($\sigma$ is defined as the ratio between Poynting flux and matter flux) flow with a typical mini-jet Lorentz factor $\gamma \sim \sqrt{1+\sigma}$, and synchrotron radiation of the particles accelerated from these mini-jets will power fast-varying pulses in a GRB light curve \citep{lyutikov03,narayan09,Zhang-Yan11}.} A well-developed model in this regime is the internal-collision-induced magnetic reconnection and turbulence (ICMART) model \citep{Zhang-Yan11}, which envisages an emission radius $R_{\rm ICMART}\sim 10^{15}-10^{16}$ cm. 

The extremely bright GRB 230307A 
was {\yisx observed by Fermi \citep{Fermi23}} and the Gravitational wave high-energy Electromagnetic Counterpart All-sky Monitor (GECAM) \citep{Xiong23} with trigger time of 15:44:06.650 UT on 7 March 2023. The Lobster Eye Imager for Astronomy (LEIA, the pathfinder of the Einstein Probe mission) also observed its prompt emission in the soft X-ray band \citep{Sunhui23}. While the burst's duration (measured as $T_{90}$) of about 41 seconds aligns with the category of long-duration GRBs, the association of a kilonova \citep{Levan23} and its unique properties in the prompt emission strongly suggest its origin from a compact binary merger event \citep{Sunhui23}. As will be demonstrated in this article, intriguing features were discovered in 
the multi-band light curves of this burst. This provides us with unique evidence that the ICMART-like mechanism was at work in the prompt emission of GRB 230307A. 

The paper is organized as follows: In the section {\bf Observation}, we will describe the observation details of GRB 230307A; then in the following section {\bf Analysis of the multi-band light curves}, we present the analysis of the multi-band data of the burst, and exhibit the aforementioned features; in the section {\bf Physical implication}, we discuss in detail why these observational features can shed light on the prompt emission mechanism. 
\section{Observation}
GRB 230307A triggered GECAM-B in real-time at 15:44:06.650 UT on 7 March 2023 (denoted as T$_0$) and also detected by GECAM-C and other gamma-ray monitors (e.g. Fermi/GBM \citep{Fermi23}, STIX \citep{STIX23}), while GECAM-A was offline at that time\footnote{GECAM is a dedicated all-sky gamma-ray monitor constellation funded by the Chinese Academy of Sciences, and now consists of three telescopes, i.e. GECAM-A and GECAM-B \citep{Li2021RDTM} micro-satellties launched together on December 10, 2020, and GECAM-C (also called High Energy Burst Searcher, HEBS) \citep{Zhang2023NIMA} onboard SATech-01 experimental satellite launched on July 27, 2022. There are two kinds of detectors in each GECAM telescope: Gamma-Ray Detectors (GRDs) and Charged Particle Detectors (CPDs). GRDs are the main detector of GECAM, each of which is composed of a scintillator and an array of SiPMs. There are 25 GRDs onboard each of GECAM-A and GECAM-B, and 12 GRDs onboard GECAM-C. {\revision GECAM is designed to have the highest time resolution (0.1 $\mu$ s) among all GRB detectors ever flown \citep{Xiao2022GECAM_time_calibrations}, making it one of the most suitable detectors for conducting the temporal research about GRBs.}}. The real-time alert data was transmitted instantly with the Global Short Message Communication of Beidou satellite navigation system \citep{Zhao2021arxiv,BDS2023IEEE}, and processed by the automatic pipeline of GECAM\cite{Huang2023arxiv}, based on which the extreme brightness of GRB 230307A was reported by GECAM to the community \citep{Xiong23}, initiating many multi-wavelength follow-up observations \citep{goodman2023GCN,JWST2023GCN1}. {\revision GRB 230307A is specially interesting because of its apparent single FRED shape in the overall light curve, and its high fluence and variability.}

Thanks to the dedicated design of instrument \citep{Zhang2023NIMA,liu2021RDTM}, neither GECAM-B nor GECAM-C suffered from data saturation during the whole burst of GRB 230307A despite of its extreme brightness \citep{Sunhui23}. High quality of GECAM data allow us to accurately measure the temporal and spectral properties of GRB 230307A. GRD04 of GECAM-B and GRD01 of GECAM-C are selected for the analysis of light curves because of their smallest incident angle to the direction of GRB 230307A. These two detectors both operate in two readout channels: high gain (HG) and low gain (LG), which are independent in terms of data processing, transmission, and dead-time. At the time of this burst, GECAM-C GRDs have a lower energy detection threshold of about 6 keV (owing to less radiation damage on SiPM) while GECAM-B GRDs have a relatively higher energy detection threshold of about 30 keV. For GRD04 of GECAM-B, the energy range of HG channel data are used from about 30 keV to 300 keV while the energy range of LG channel data are used from about 300 keV to 1000 keV. For GRD01 of GECAM-C, only HG channel data are used with the energy range from 6 keV to 30 keV. Though the response of GRD01 of GECAM-C for 6-15 keV and GRD04 of GECAM-B for 300-700 keV is affected by the electronics, this does not have any effect on the analysis of light curves. The background of GECAM-B is estimated by fitting the data from $T_0$-50 s to $T_0$-5 s and $T_0$+160 s to $T_0$+200 s with the first order polynomials. The background of GECAM-C is estimated by fitting the data from $T_0$-20 s to $T_0$-1 s and $T_0$+170 s to $T_0$+600 s with a combination of the first and second-order exponential polynomials \citep{Sunhui23}. {\revision The observed light curves are not directly used to identify features of the broad pulse and fast pulses. Those features are identified through
jointly fitting the background and the signals using various functional forms to describe pulses (see in the following section).}

{\revision We have also cross-checked the GECAM data with that of Fermi/GBM, which has suffered from pile-up effects, which severely changed the shape of the light curve around the peak. For other part of the GBM data that did not suffer from pile-up, we performed the same analysis, including temporal analysis and spectral analysis. The results show well consistency with the GECAM results. We therefore do not include the GBM data in our following analysis. }

\section{Analysis of the multiband light curves}
\subsection{The Norris-like single pulse overall profile}
\begin{figure}
    \centering
    \includegraphics[width=\linewidth]{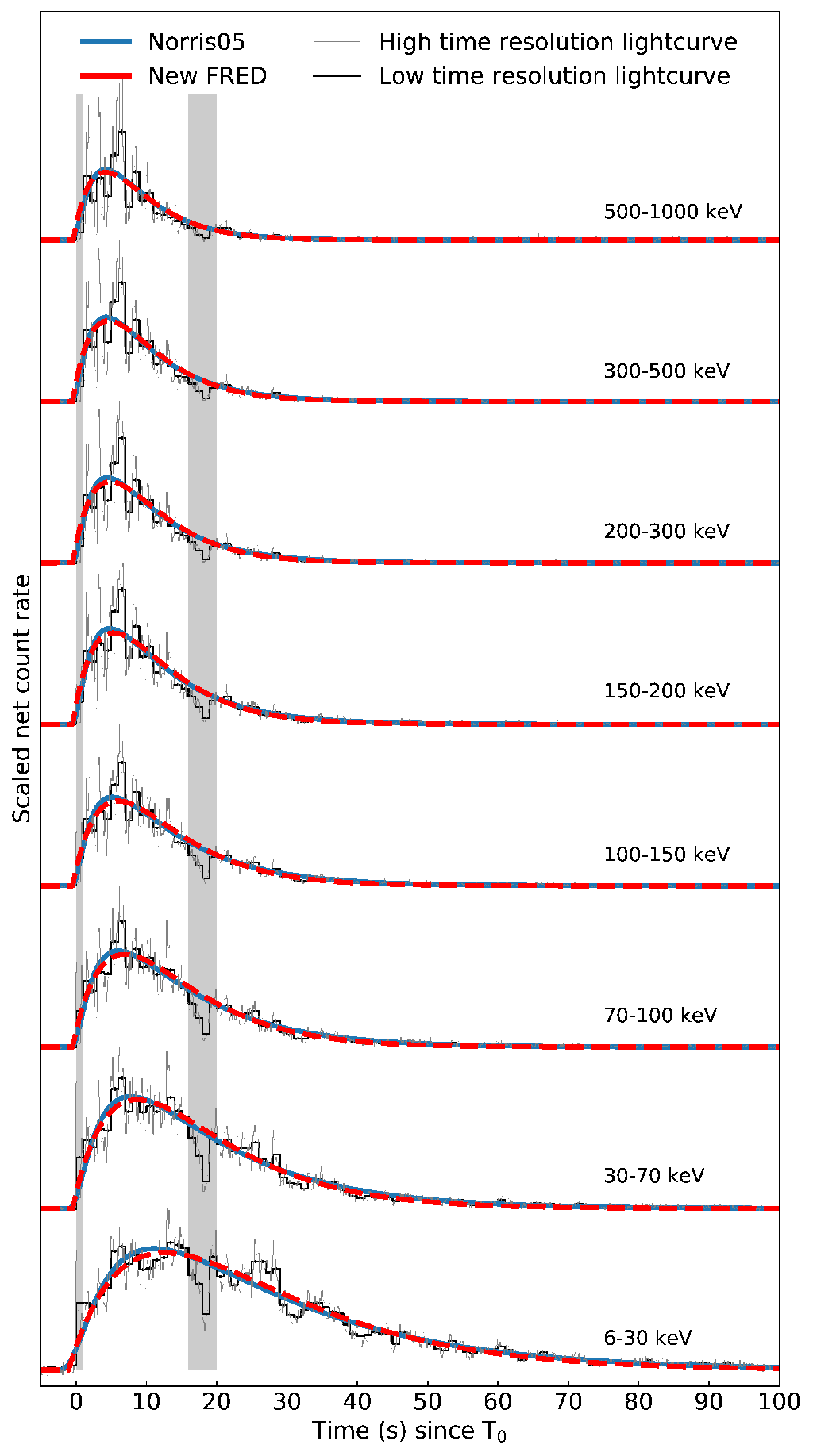}
    \caption{{\bf The multi-band light curves.} The black histograms are 1\,s binned net count rates (with background subtracted), whose profile shows a single FRED shape; The gray thin histograms are 0.1\,s binned net count rates (with background subtracted); The blue solid curves are best-fit FRED model with the Norris05 formulation, while the blue dashed curves are the best-fit with our new FRED formulation. Gray shadowed regions (the precursor and the dip) are ignored in the fitting. All error bars represent 1$\sigma$ uncertainties of the net count rates.}
    \label{fig:1}
\end{figure}

The unsaturated data record by GECAM enable us to accurately characterize the temporal properties of GRB 230307A. {\yisx The high time-resolution light curves show many rapidly varying structures in the prompt emission (see thin grey lines in Figure \ref{fig:1}). Intriguingly, by comparing the light curves in different energy bands\footnote{namely, 6-30 keV, where the data are from GECAM-C; 30-70 keV, 70-100 keV, 100-150 keV, 150-200 keV, 200-300 keV, 300-500 keV and 500-1000 keV, where the data are from GECAM-B}, we found that the light curves also have slow varying trends, which are slower at lower energy bands (this can be better seen with lower time-resolution light curves. See the binned light curves in Figure \ref{fig:1}).} 
{\yisx In order to quantitatively study the slow trends, we attempt to fit the low time-resolution light curves with a conventional parameterisation of the FRED (Fast Rise Exponential Decay) profile} \citep{Norris05}: 
\begin{equation}
    L(t)\propto \frac{1}{\exp(\frac{\tau_{\rm{r}}}{t-t_s}+\frac{t-t_s}{\tau_{\rm{d}}})}.
    \label{eq:Norris05}
\end{equation}
In the above formulation, $t_{\rm{s}}$ is the starting instance of the {\yisx profile}, $\tau_{\rm{r}}$ and $\tau_{\rm{d}}$ are the rising and decaying time scales, respectively. In the fitting process, we assume that the net count (with background model subtracted) in each bin follows a Poisson probability distribution, with the expected value equal to the parameterized FRED profile. The fittings are done with the maximum likelihood method, which can be expressed as ${\rm ln}\mathcal{L}=\sum_i [{\rm MODEL}_i-{\rm DATA}_i{\rm ln}({\rm MODEL}_i)]$, where the subscript \textit{i} runs over all time bins from -4\,s to 100\,s (excluding 0\,s to 1\,s and 16\,s to 20\,s) and the part used to estimate the background model. The posteriors of the fitted parameters are found with a Monte Carlo Markov Chain (MCMC) method.

The fitting results are listed in Table \ref{tab:total_lc_fit_pars}. The best-fit Norris FRED profiles are plotted with multi-band light curves in Figure 1. The peak time $t_{\rm{p}}$ and the width $w$ of the {\yisx} are therefore defined as: 
$t_{\rm{p}}=t_{\rm{s}}+\sqrt{\tau_{\rm{r}}\tau_{\rm{d}}}$ and $w=\tau_{\rm{r}}+\tau_{\rm{d}}$. It can be seen in Figure {fig:1.5} that there is a clear energy dependence in both $w$ and $\tilde{t}_{\rm{p}}\equiv t_{\rm{p}}-t_{\rm{s}}$: {\yisx In the energy range from 6 keV to 300 keV, the $w-E$ and $\tilde{t}_{\rm{p}}-E$ relations are both in a power law function, with the power indices {\revision-0.38$^{+0.05}_{-0.06}$ and -0.37$^{+0.05}_{-0.06}$}, respectively. Above $\sim300$ keV, the energy dependence on both $w$ and $\tilde{t}_{\rm{p}}$ saturate. Such a profile-energy dependence was commonly found in single GRB pulses \citep{Norris05,Liang06,Hakkila08,Peng12}.}
\begin{figure}
    \centering
    \includegraphics[width= 8.5 cm ]{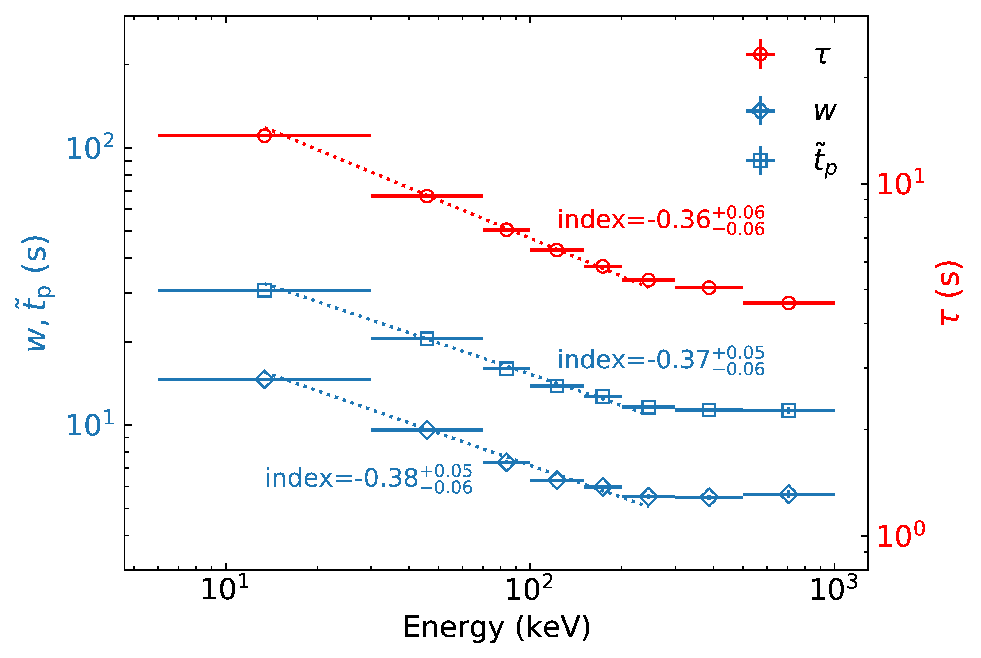}
    \caption{{\bf The fitted FRED formulation parameters as a function of energy.} The vertical error bars indicate the 1$\sigma$ uncertainties of the fitted parameters, while the horizontal error bars indicate the ranges of the energy bins.}
    \label{fig:1.5}
\end{figure}
\subsection{The self-similarity of the overall profile}
We would like to further explore the profile-energy dependence. As we found above, the peak time and the width of the profile vary with energy with the same power law index. As a consequence, the overall profile of the light curve (the broad pulse) in each energy band is a time-stretch copy of that in the other bands (we follow the terminology of Ref. \cite{Norris05} to call it a ``self similar profile"). In order to further demonstrate the self-similar feature of the slow varying profile, we smooth the light curves in a non-parametrical way, and manifest their shape identity after energy-dependent re-scaling in time.

\begin{figure}
    \centering
    \includegraphics[width=8 cm]{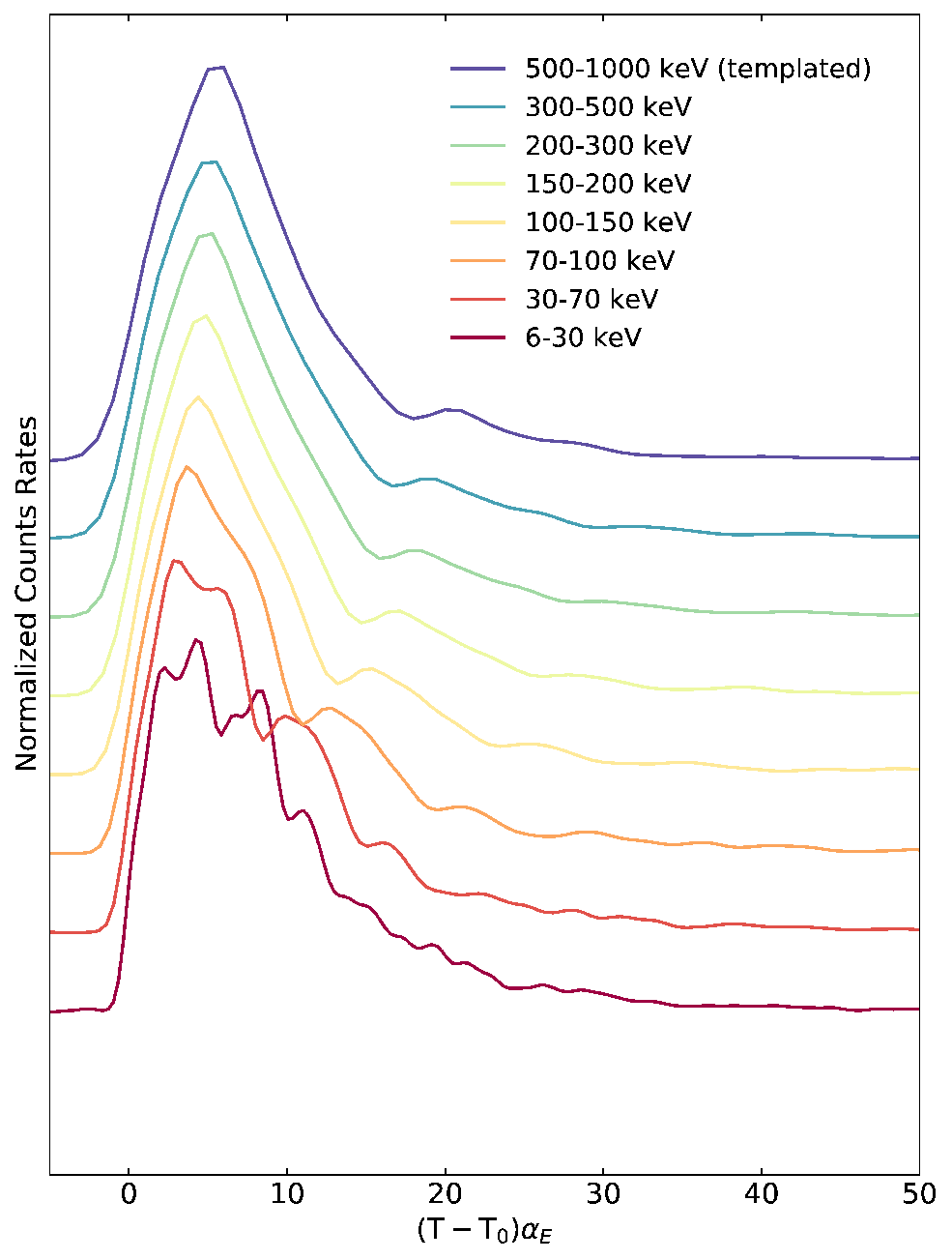}
    \caption{\bf The smoothed and time-rescaled GECAM multiband light curves of GRB 230307A}
    \label{fig:2}
\end{figure}

\begin{figure}
    \centering
    \includegraphics[width=8 cm]{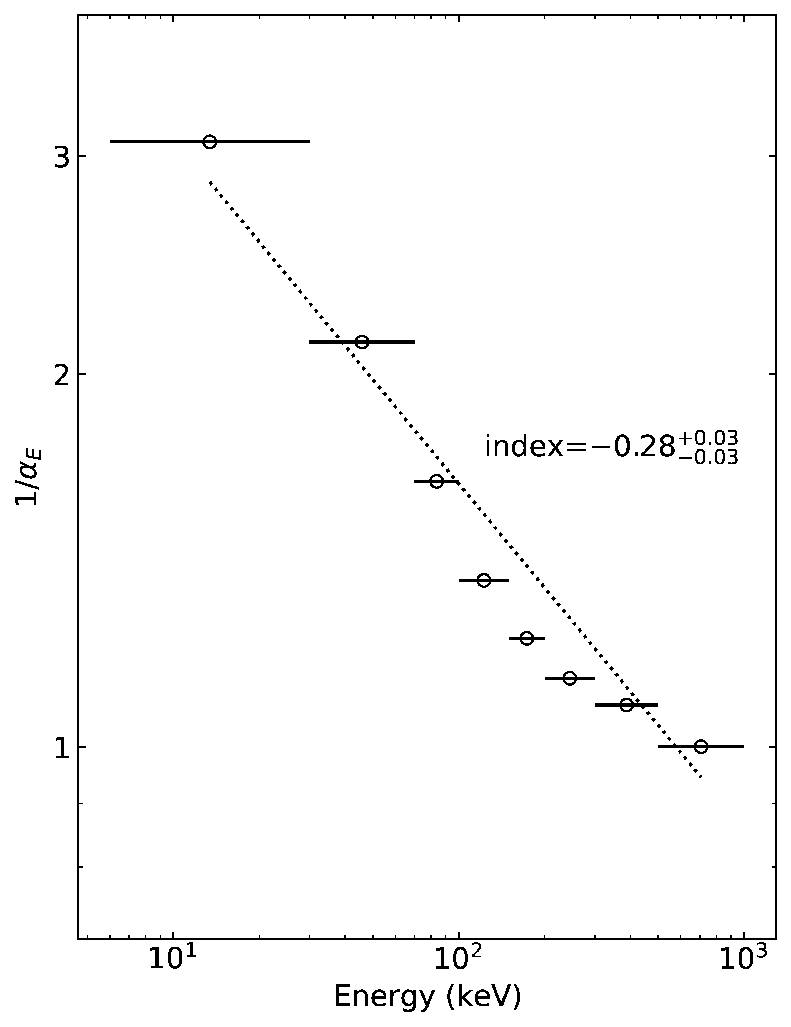}
    \caption{{\bf The scaling factor of the Gaussian smoothed light curve as a function of energy}. The vertical error bars indicate the 1$\sigma$ uncertainties of the fitted parameters, while the horizontal error bars indicate the ranges of the energy bins.}
    \label{fig:3}
\end{figure}
The smoothed light curves are convolution between a Gaussian kernel and the original light curves. The sigma of the Gaussian kernel is 1.5\,s. It can be obviously shown that, the shape of the profiles in different energy bands can be "stretched" in time domain into an identical shape with an energy dependent scaling factor. We use the smoothed light curve in the 500-1000 keV band as the template, and we fit the template to light curves in other bands, which are scaled by multiplying scaling factors $\alpha_{\rm{E}}$ to the their time argument $T-T_0$. We stack all the re-scaled and smoothed light curves in Figure \ref{fig:2}. It is intuitively shown that the self-similarity of the broad profile, and this conclusion is FRED formulation independent. The scaling factor as a function of energy is plotted in the Figure \ref{fig:3}, which again shows a power law dependence. 
\begin{table*}
\centering
\scriptsize
\caption{\textbf{Fitting results for the light curves of GRB 230307A.} All errors represent the 1$\sigma$ uncertainties.}
\label{tab:total_lc_fit_pars}
\begin{tabular}{c|cccc|ccc}
\hline
 & \multicolumn{4}{c}{Norris05} & \multicolumn{3}{|c}{New FRED}\\
\cline{2-8}
Energy range & $\tau_{\rm r}$ & $\tau_{\rm d}$ & norm & t$_s$ & $\tau$ & norm & t$_s$  \\ 
(keV) & (s) & (s) & (counts$\cdot$s$^{-1}$) & (s) & (s) & (counts$\cdot$s$^{-1}$) & (s) \\ 
\hline 
6-30 & $10.78^{+0.31}_{-0.30}$ & $19.87^{+0.13}_{-0.13}$ & $2.49^{+0.06}_{-0.06}\times10^{4}$ & $-3.53^{+0.11}_{-0.11}$ & $13.68^{+0.04}_{-0.04}$ & $1.50^{+0.01}_{-0.01}\times10^{4}$ & $-1.08^{+0.04}_{-0.04}$\\  
30-70 & $6.65^{+0.20}_{-0.19}$ & $13.85^{+0.10}_{-0.10}$ & $2.04^{+0.05}_{-0.05}\times10^{4}$ & $-1.85^{+0.06}_{-0.06}$ & $9.22^{+0.03}_{-0.03}$ & $1.35^{+0.01}_{-0.01}\times10^{4}$ & $-5.00^{+0.08}_{-0.07}\times10^{-1}$\\  
70-100 & $4.84^{+0.17}_{-0.17}$ & $11.14^{+0.09}_{-0.09}$ & $1.17^{+0.03}_{-0.03}\times10^{4}$ & $-1.31^{+0.06}_{-0.06}$ & $7.40^{+0.03}_{-0.03}$ & $8.20^{+0.05}_{-0.05}\times10^{3}$ & $-4.94^{+0.14}_{-0.07}\times10^{-1}$\\  
100-150 & $4.09^{+0.13}_{-0.13}$ & $9.74^{+0.07}_{-0.07}$ & $1.44^{+0.04}_{-0.04}\times10^{4}$ & $-1.15^{+0.05}_{-0.05}$ & $6.48^{+0.03}_{-0.03}$ & $1.02^{+0.01}_{-0.01}\times10^{4}$ & $-4.96^{+0.09}_{-0.04}\times10^{-1}$\\  
150-200 & $4.25^{+0.17}_{-0.17}$ & $8.43^{+0.08}_{-0.08}$ & $1.10^{+0.04}_{-0.04}\times10^{4}$ & $-1.19^{+0.06}_{-0.06}$ & $5.82^{+0.03}_{-0.03}$ & $6.88^{+0.05}_{-0.05}\times10^{3}$ & $-4.90^{+0.20}_{-0.10}\times10^{-1}$\\  
200-300 & $4.00^{+0.18}_{-0.18}$ & $7.59^{+0.08}_{-0.08}$ & $1.07^{+0.04}_{-0.04}\times10^{4}$ & $-1.07^{+0.06}_{-0.06}$ & $5.32^{+0.03}_{-0.03}$ & $6.49^{+0.05}_{-0.05}\times10^{3}$ & $-4.83^{+0.28}_{-0.14}\times10^{-1}$\\  
300-500 & $4.25^{+0.19}_{-0.18}$ & $7.08^{+0.08}_{-0.08}$ & $1.23^{+0.05}_{-0.05}\times10^{4}$ & $-1.20^{+0.06}_{-0.05}$ & $5.06^{+0.03}_{-0.03}$ & $6.77^{+0.05}_{-0.05}\times10^{3}$ & $-4.86^{+0.26}_{-0.13}\times10^{-1}$\\  
500-1000 & $5.38^{+0.42}_{-0.39}$ & $5.89^{+0.12}_{-0.12}$ & $8.00^{+0.76}_{-0.67}\times10^{3}$ & $-1.41^{+0.10}_{-0.10}$ & $4.58^{+0.05}_{-0.05}$ & $3.09^{+0.04}_{-0.04}\times10^{3}$ & $-4.00^{+0.90}_{-0.71}\times10^{-1}$\\  
\hline
\end{tabular}
\end{table*}

The ``self-similarity" inspired us to propose another formulation of FRED profile, with one less parameter than that of Norris05: 
\begin{equation}
    L(t)\propto \frac{t-t_{\rm{s}}}{\tau}\exp(-\frac{t-t_{\rm{s}}}{\tau}).
    \label{eq:YiFred}
\end{equation}

We fit the multi-band light curves again with the new FRED formulation. The results of fitting are listed in Table 1.
The sole time scale parameter $\tau$ shows a similar energy dependence on energy, 
{\revision$\tau\propto E^{-0.36^{+0.06}_{-0.06}}$}, from 6 to 300 keV, and shows a hint of shallower energy dependence above $\sim300$ keV (Figure \ref{fig:1.5}). In the new formulation, one can define peak time as $t_{\rm p}=t_s+\tau_E$, where we denote the $\tau$ in the $E$ channel with $\tau_E$. The width of the FRED profile also scales as $\tau_E$. Therefore, the found $\tau-E$ dependence can naturally result in the $w-E$ and $\tilde{t}_{\rm{p}}-E$ relations. 

\subsection{The fast components}
\begin{figure}
    \centering
    \includegraphics[width=\linewidth]{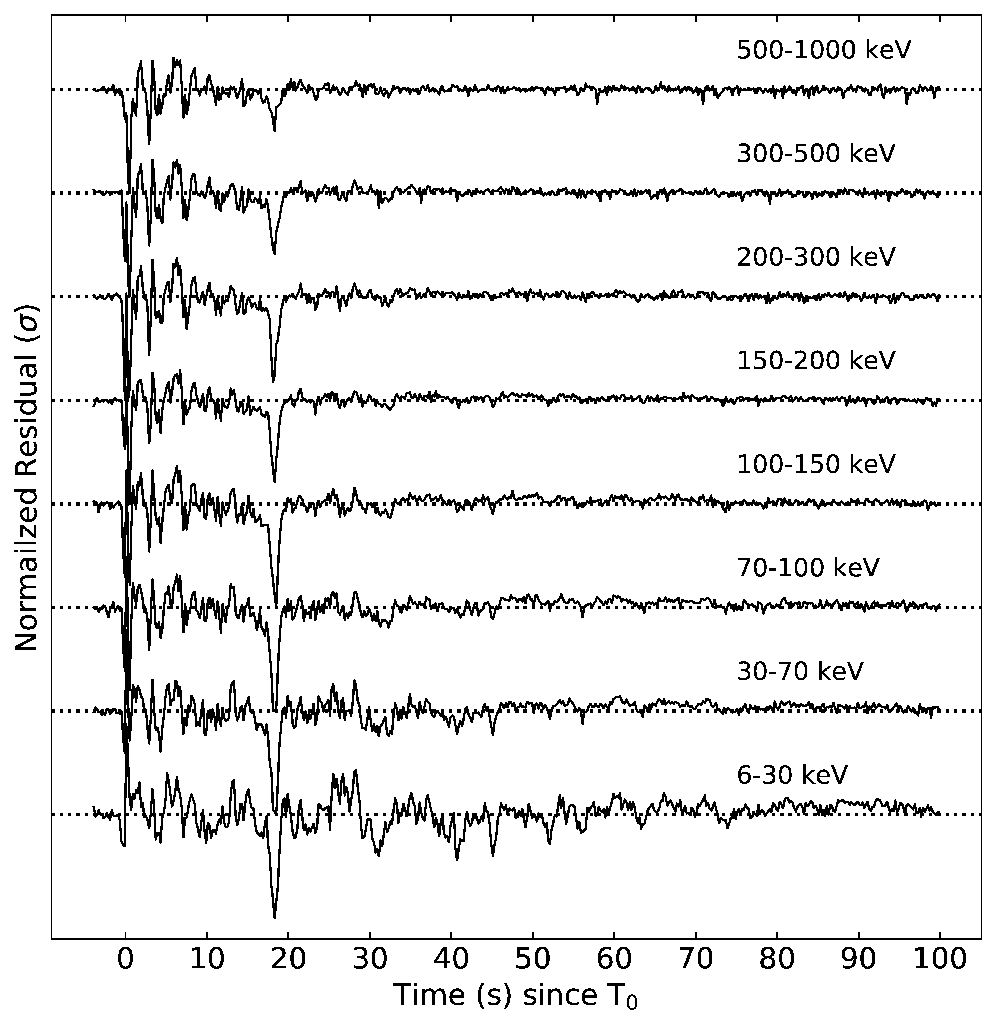}
    \caption{The residuals of the net light curves after fitting with the new FRED formulation}
    \label{fig:4}
\end{figure}

{\yisx We subtract the slow trend described by equation \ref{eq:YiFred} with the best-fit parameters, to study the fast varying structures.} From the residuals of the light curves (Figure \ref{fig:4}), one can see that there are many rapidly-varying short pulses. One noticeable feature is that even the broad pulse has a clear energy dependence, the 
short time spikes and dips appear to align at the same instances across different energy bands.

\begin{figure}
    \centering
    \includegraphics[width=\linewidth]{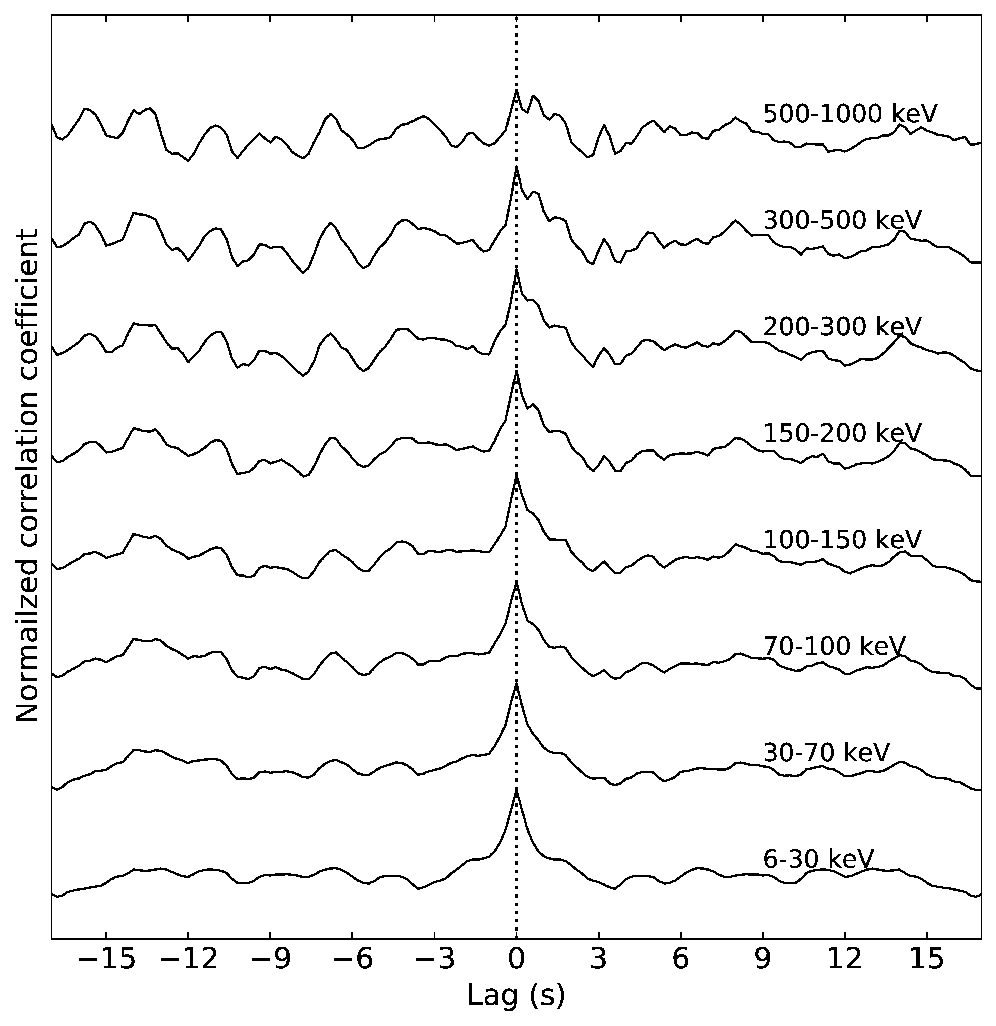}
    \caption{{\bf The cross-correlation between the multi-band residuals and that in the 6-30 keV channel.} The vertical dotted line indicates the zero lag time.}
    \label{fig:5}
\end{figure}

We further demonstrate this alignment of the fast temporal features by cross-correlations among multi-band light curve residuals. As shown in the Figure \ref{fig:5}, the peaks are perfectly aligned across the full energy band. This immediately excludes the attempt to explain the self-similar stretchable feature of the multi-band light curves as relativistic time-dilation, as a time-dilation effect would stretch both slow and fast varying pulses together. 
\begin{figure}
    \centering
    \includegraphics[width=\linewidth]{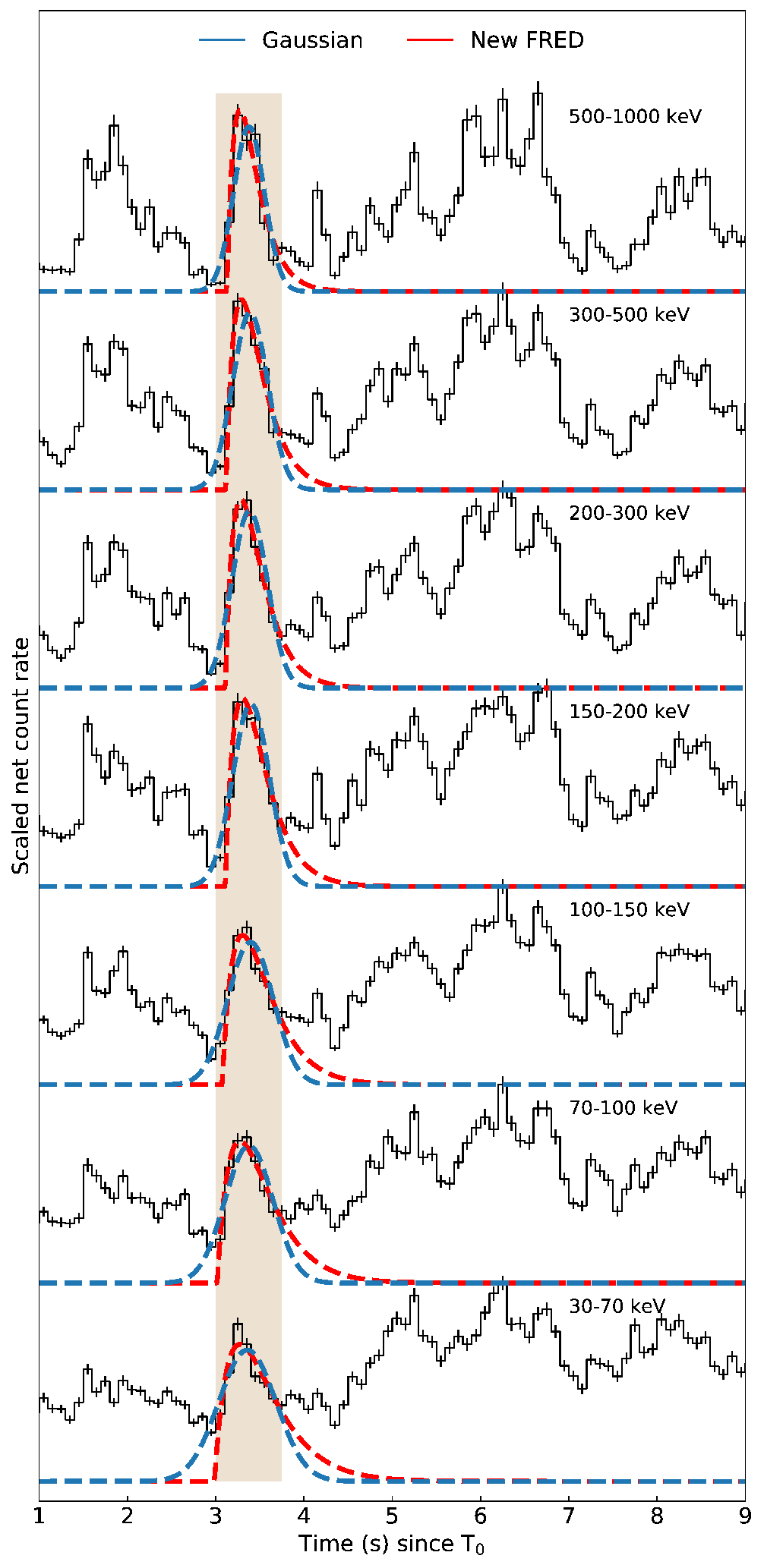}
    \caption{{\bf Pulse profile fitting to the fast varying temporal structure at $\sim3.5$\,s.} Red and blue dashed lines correspond to new FRED formulation and Gaussian profile respectively. The shadowed region indicates the range of data involved in the fitting.}
    \label{fig:6}
\end{figure}
\begin{figure}
    \centering
    \includegraphics[width=\linewidth]{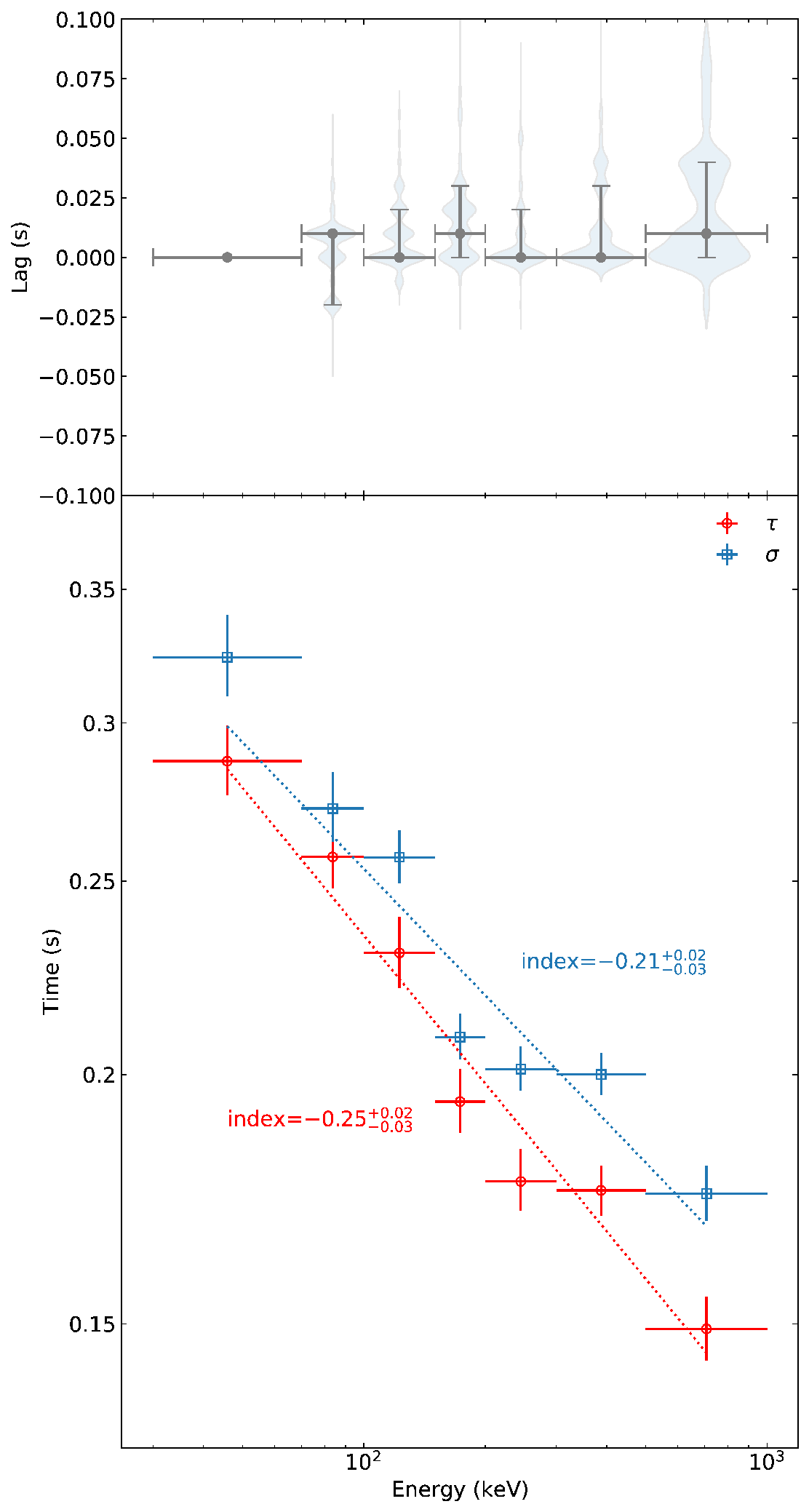}
    \caption{{\bf Upper panel:} Cross-correlation between the light curves of small pulse in multi-bands as functions of the energy. {\bf Lower panel: }The fitted width ($\tau$ for new FRED and $\sigma$ for Gaussian profile) of pulse at $\sim3.5$\,s as function of energy.}
    \label{fig:7}
\end{figure}

Here we further demonstrate that some of the prominent fast structures can be identified as small pulses without an underlying slow component. For instance, we show that the spike at $\sim3.5$\,s can be fitted with a pulse profile. We employ our new FRED formulation to fit the pulse from 3\,s to 3.75\,s, and find the typical width-$E$ relation ($\tau-E$ in the new FRED formulation) in GRB pulses. In order to show this conclusion is independent on the pulse profile formulation, we use Gaussian function as an alternative profile to fit the spike, and obtain the same width-$E$ relation ($\sigma-E$  with the Gaussian pulse formulation. See the lower panel of Figure \ref{fig:7}. It can be seen in the Figure \ref{fig:6} that {\yisxapj the FRED profile can also be used to fit the individual pulse, with better representation of its asymmetry}. The fitted parameters are list in Table \ref{tab:fast_fit_pars}. It is also interesting to observe that this small pulse does not show spectrum lag of its peak time as the broad pulse. We demonstrate this by cross-correlating between the small pulses in multi-bands and that in 30-70 keV channel. The lag time together with their uncertainties as a function of energy is shown in the upper panel of Figure \ref{fig:7}, where all lag times are almost zero, and we can clearly conclude that the lag time does not depend on energy like the broad pulse.

\begin{table*}
\centering
\scriptsize
\caption{\textbf{Light curves fitting results for the fast varying temporal structure of light curves at about 3.5 s .} All errors represent the 1$\sigma$ uncertainties.}
\label{tab:fast_fit_pars}
\begin{tabular}{c|ccc|ccc}
\hline
 & \multicolumn{3}{c}{Gaussian} & \multicolumn{3}{|c}{New FRED}\\
\cline{2-7}
Energy range & $\mu$ & $\sigma$ & norm & $\tau$ & norm & t$_s$ \\ 
(keV) & (s) & (s) & (counts$\cdot$s$^{-1}$) & (s) & (counts$\cdot$s$^{-1}$) & (s) \\ 
\hline
30-70 & $3.36^{+0.01}_{-0.01}$ & $3.23^{+0.16}_{-0.14}\times10^{-1}$ & $4.87^{+0.13}_{-0.13}\times10^{3}$ & $2.87^{+0.12}_{-0.11}\times10^{-1}$ & $1.38^{+0.04}_{-0.04}\times10^{4}$ & $2.99^{+0.01}_{-0.01}$\\  
70-100 & $3.38^{+0.01}_{-0.01}$ & $2.72^{+0.12}_{-0.10}\times10^{-1}$ & $3.63^{+0.11}_{-0.11}\times10^{3}$ & $2.57^{+0.10}_{-0.09}\times10^{-1}$ & $1.02^{+0.03}_{-0.03}\times10^{4}$ & $3.02^{+0.004}_{-0.005}$\\  
100-150 & $3.39^{+0.01}_{-0.01}$ & $2.57^{+0.08}_{-0.07}\times10^{-1}$ & $5.16^{+0.13}_{-0.13}\times10^{3}$ & $2.30^{+0.10}_{-0.09}\times10^{-1}$ & $1.47^{+0.04}_{-0.04}\times10^{4}$ & $3.07^{+0.01}_{-0.01}$\\  
150-200 & $3.40^{+0.01}_{-0.01}$ & $2.09^{+0.06}_{-0.05}\times10^{-1}$ & $4.11^{+0.12}_{-0.12}\times10^{3}$ & $1.94^{+0.08}_{-0.07}\times10^{-1}$ & $1.15^{+0.04}_{-0.04}\times10^{4}$ & $3.11^{+0.005}_{-0.01}$\\  
200-300 & $3.38^{+0.01}_{-0.01}$ & $2.01^{+0.05}_{-0.05}\times10^{-1}$ & $4.23^{+0.13}_{-0.12}\times10^{3}$ & $1.77^{+0.06}_{-0.06}\times10^{-1}$ & $1.22^{+0.04}_{-0.04}\times10^{4}$ & $3.11^{+0.005}_{-0.01}$\\  
300-500 & $3.39^{+0.005}_{-0.01}$ & $2.00^{+0.05}_{-0.05}\times10^{-1}$ & $4.45^{+0.13}_{-0.13}\times10^{3}$ & $1.75^{+0.05}_{-0.05}\times10^{-1}$ & $1.30^{+0.04}_{-0.04}\times10^{4}$ & $3.12^{+0.003}_{-0.003}$\\  
500-1000 & $3.38^{+0.01}_{-0.01}$ & $1.74^{+0.06}_{-0.05}\times10^{-1}$ & $2.25^{+0.10}_{-0.09}\times10^{3}$ & $1.49^{+0.06}_{-0.05}\times10^{-1}$ & $6.72^{+0.31}_{-0.29}\times10^{3}$ & $3.13^{+0.003}_{-0.003}$\\  
\hline
\end{tabular}
\end{table*}
\subsection{The dip at around 18s}
\begin{figure}
    \centering
    \includegraphics[width=\linewidth]{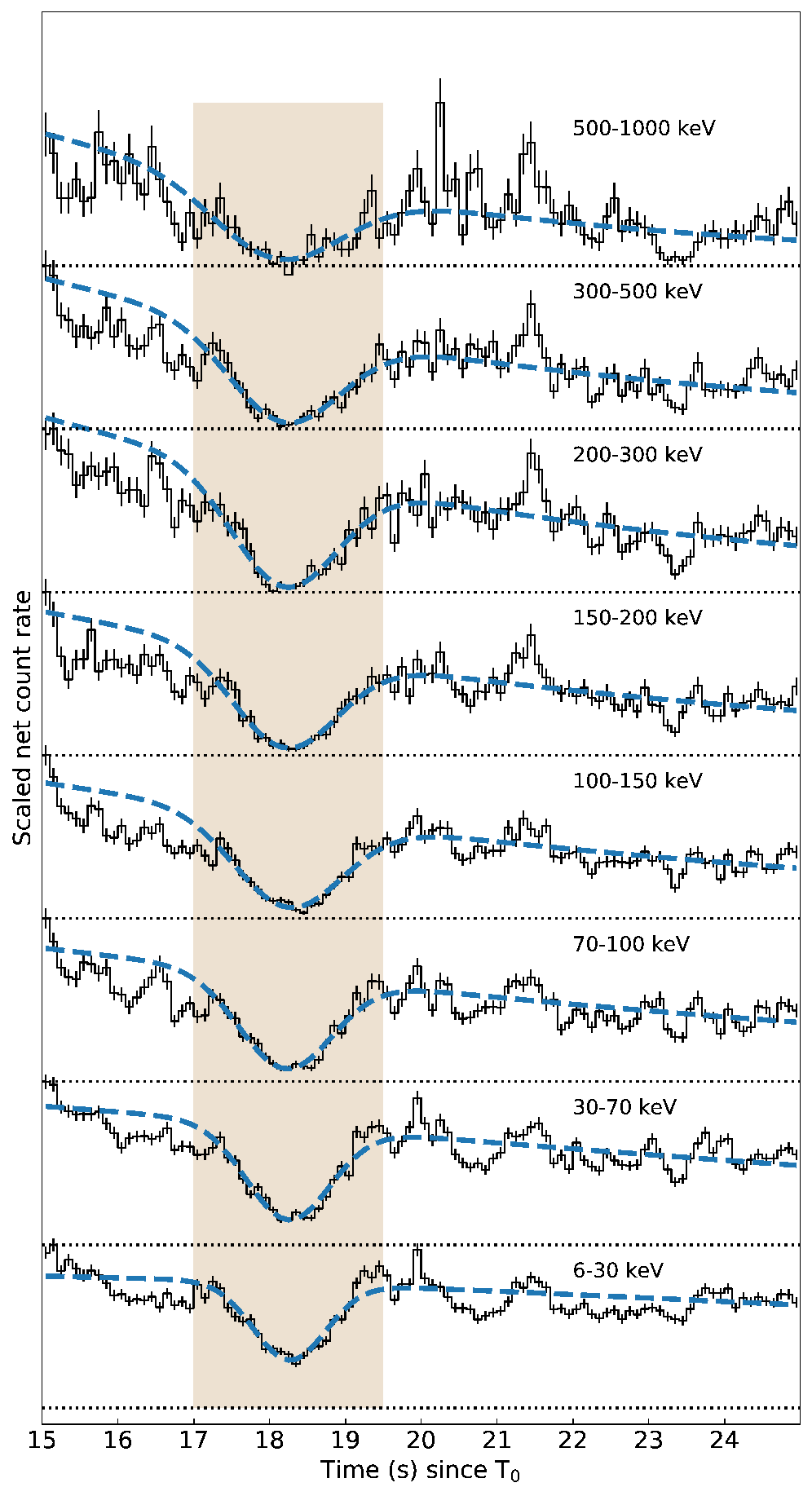}
    \caption{Fitting the gap with a FRED profile and a negative Gaussian function. The shadowed region indicates the range of data involved in the fitting}
    \label{fig:7.5}
\end{figure}

\begin{figure}
    \centering
    \includegraphics[width=\linewidth]{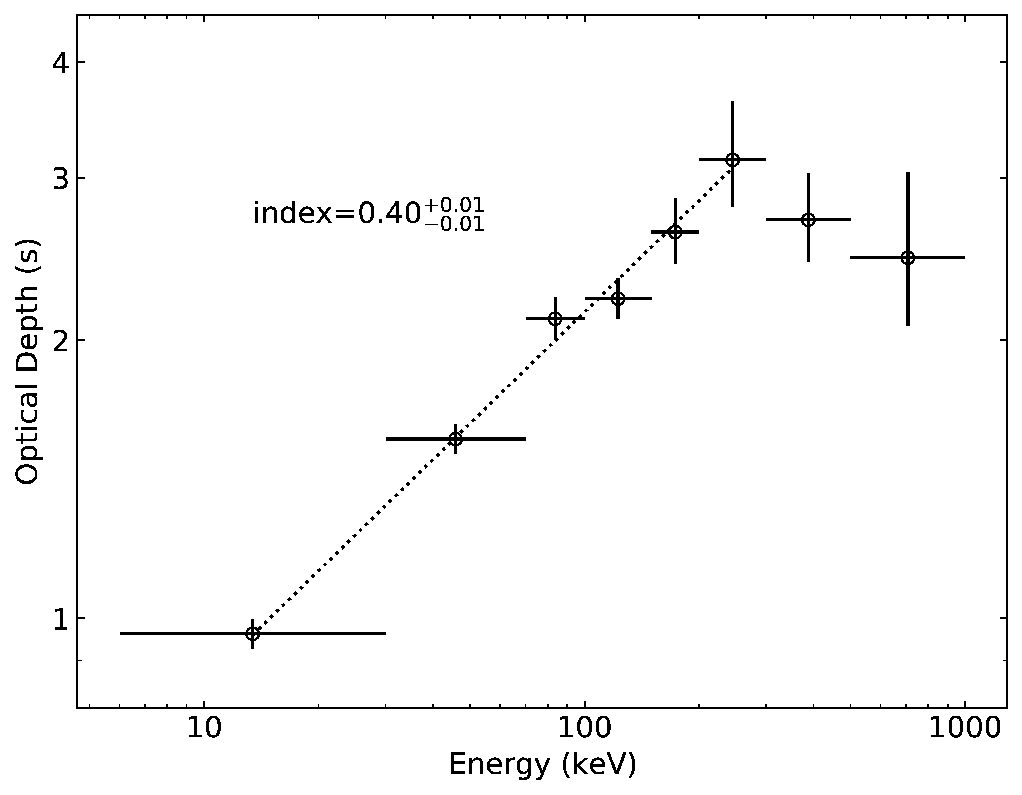}
    \caption{The effective optical depth defined with equation \ref{eq:Optdepth}, as a function of energy.}
    \label{fig:8}
\end{figure}

The dip at $\sim18$\,s is another noticeable feature of the multi-band light curves. A natural attempt is to attribute such a dip to some sort of absorption or geometrical blocking. We define the effective ``optical depth" $\tau_{\rm{op},E}$ in different energy bands as:
\begin{equation}
    \exp(-\tau_{\rm{op},E})=\frac{C_{\rm{dip},E}}{C_{\rm{slow},E}},
    \label{eq:Optdepth}
\end{equation}
where $C_{\rm dip}$ is the net count rate at the bottom of the dip, which is found with a negative Gaussian fitting superposed on the FRED of broad pulse fitting from 17\,s to 19.5\,s (Figure \ref{fig:7.5}); and $C_{\rm slow}$ is the net count rate of the broad feature. The results of the fitting with a negative Gaussian is tabulate in Table \ref{tab:optical_depth_pars}. It is intriguing to find that there is a power law energy dependence of the effective optical depth, and the best fit power law index is $\sim0.4$ (Figure \ref{fig:8}). This energy dependence of the optical depth challenges the absorption picture, for there is no known absorption mechanism whose cross section proportional to energy to the order of $\sim0.4$ {\revision with a break at about 300\,keV}. {\revision Beside, if we were to explain the dip as absorption, its depth and transience require that the absorbing object has a angular size comparable to that of the emission region, and a very high transverse angular velocity. That will either result in a superluminal transverse velocity of the blocking object, or highly coincidental absorption in our local.}
\begin{table*}
\centering
\scriptsize
\caption{\textbf{Light curves fitting results for the dip and optical depth.} All errors represent the 1$\sigma$ uncertainties.}
\label{tab:optical_depth_pars}
\begin{tabular}{ccccc}
\hline
Energy range & $\mu$ & $\sigma$ & $C_{{\rm slow},E}$ & Optical Depth \\ 
(keV) & (s) & (s) & (counts$\cdot$s$^{-1}$) &  \\ 
\hline
6-30 & $18.26^{+0.01}_{-0.01}$ & $4.76^{+0.14}_{-0.13}\times10^{-1}$ & $5.15^{+0.01}_{-0.01}\times10^{3}$ & $0.96^{+0.04}_{-0.03}$\\  
30-70 & $18.25^{+0.01}_{-0.01}$ & $5.51^{+0.14}_{-0.14}\times10^{-1}$ & $3.67^{+0.01}_{-0.01}\times10^{3}$ & $1.56^{+0.06}_{-0.06}$\\  
70-100 & $18.21^{+0.02}_{-0.02}$ & $6.10^{+0.19}_{-0.18}\times10^{-1}$ & $1.72^{+0.01}_{-0.01}\times10^{3}$ & $2.11^{+0.12}_{-0.11}$\\  
100-150 & $18.24^{+0.02}_{-0.02}$ & $7.07^{+0.21}_{-0.21}\times10^{-1}$ & $1.73^{+0.01}_{-0.01}\times10^{3}$ & $2.22^{+0.12}_{-0.11}$\\  
150-200 & $18.20^{+0.02}_{-0.02}$ & $6.76^{+0.26}_{-0.25}\times10^{-1}$ & $9.40^{+0.01}_{-0.01}\times10^{2}$ & $2.62^{+0.23}_{-0.20}$\\  
200-300 & $18.19^{+0.02}_{-0.02}$ & $6.87^{+0.30}_{-0.28}\times10^{-1}$ & $7.20^{+0.01}_{-0.01}\times10^{2}$ & $3.14^{+0.49}_{-0.35}$\\  
300-500 & $18.19^{+0.03}_{-0.03}$ & $7.29^{+0.35}_{-0.33}\times10^{-1}$ & $6.58^{+0.01}_{-0.01}\times10^{2}$ & $2.70^{+0.34}_{-0.27}$\\  
500-1000 & $18.09^{+0.06}_{-0.06}$ & $8.36^{+0.82}_{-0.74}\times10^{-1}$ & $2.26^{+0.01}_{-0.01}\times10^{2}$ & $2.46^{+0.59}_{-0.39}$\\  
\hline
\end{tabular}
\end{table*}

\begin{figure}
    \centering
    \includegraphics[width=\linewidth]{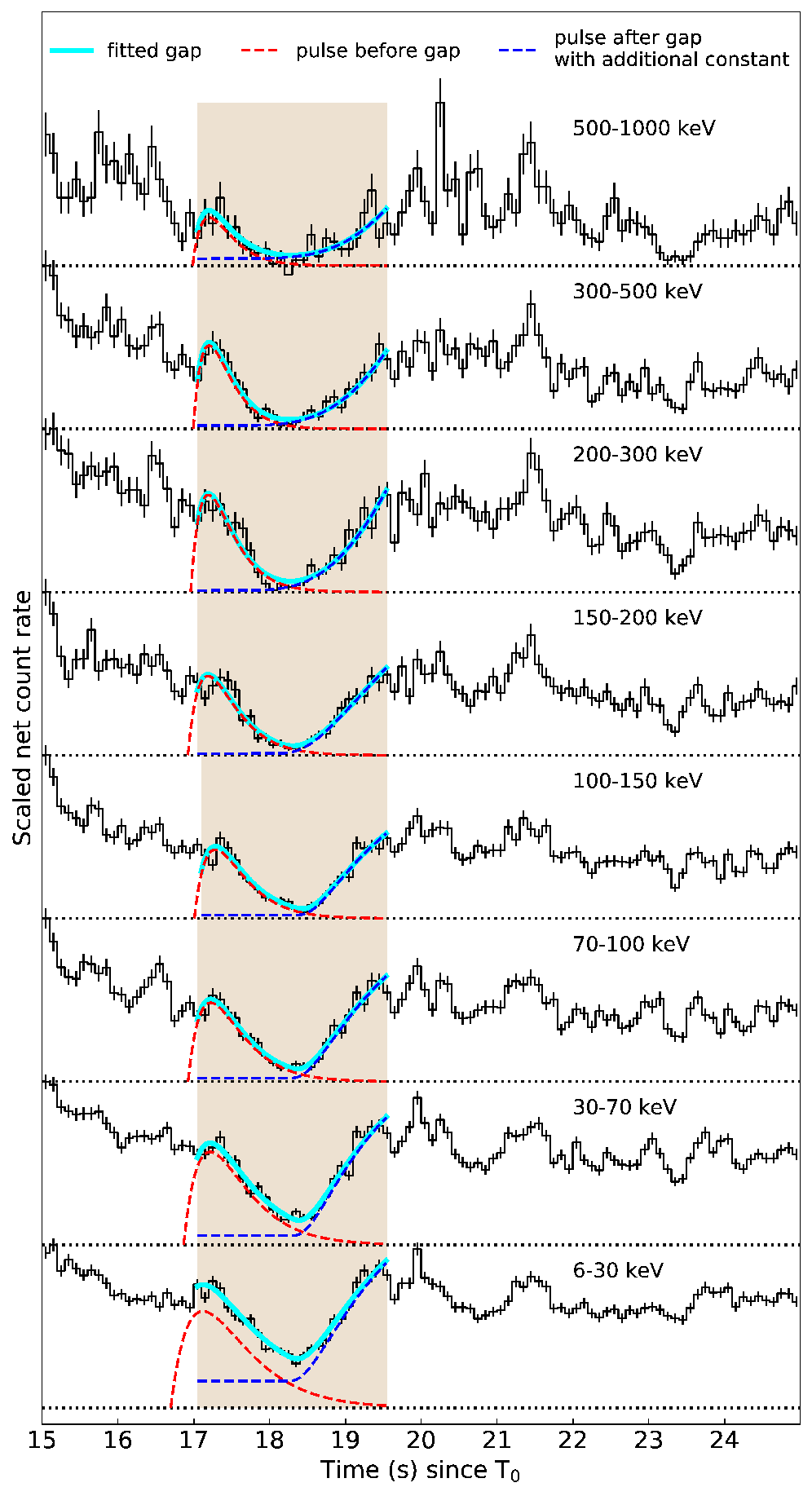}
    \caption{ {\bf Fitting the gap with two successive short pulses}. The dashed red curves indicate the earlier pulse component, and the dashed blue curves indicate the rising edge of the later pulse. Cyan curves indicating the summation of these two components. }
    \label{fig:9}
\end{figure}

\begin{figure}
    \centering
    \includegraphics[width=\linewidth]{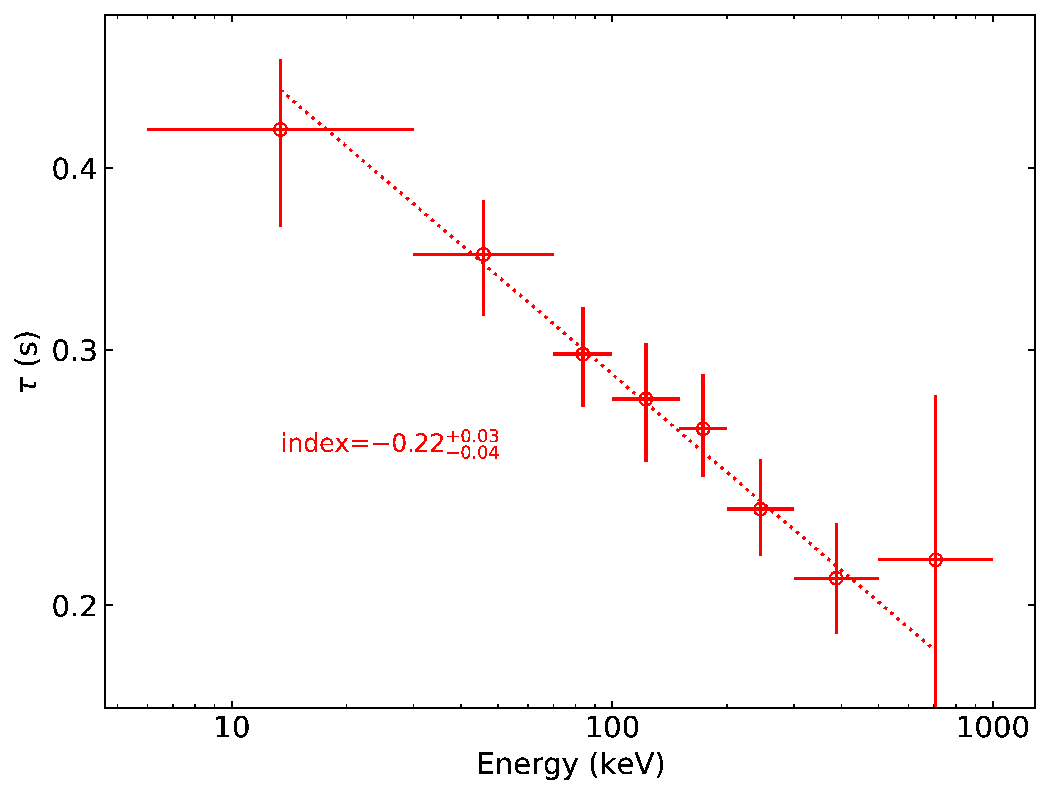}
    \caption{The time scale factor $\tau$ for new FRED of the earlier pulse component in the 18-s dip, as a function of energy}
    \label{fig:10}
\end{figure}

A reasonable explanation of the dip is the gap between two successive fast temporal structures. In order to demonstrate, we fit a pulse profile (new FRED formulation) and a rising edge of another pulse (Norris05) to the dip at each energy bands from $\sim$17\,s to 19.5\,s (Figure \ref{fig:9}). The fitted parameters are list in Table \ref{tab:gap_pars}. The time scale $\tau$ of the earlier pulses as a function of energy is plotted in Figure \ref{fig:10}, which shows clearly a typical power law energy dependence of pulse width. 

\begin{table*}
\centering
\scriptsize
\caption{\textbf{Light curves fitting results for the gap with two successive fast temporal structures.} All errors represent the 1$\sigma$ uncertainties.}
\label{tab:gap_pars}
\begin{tabular}{c|ccc|ccc|c}
\hline
 & \multicolumn{3}{c}{pulse before gap} & \multicolumn{3}{|c|}{pulse after gap} & additional constant\\
\cline{2-8}
Energy range & t$_s$ & $\tau$ & norm & t$_s$ & $\tau_{\rm r}$ & norm & constant\\
(keV) & (s) & (s) & (counts$\cdot$s$^{-1}$) & (s) & (s) & (counts$\cdot$s$^{-1}$) & (counts$\cdot$s$^{-1}$) \\ 
\hline
6-30 & $16.70^{+0.09}_{-0.09}$ & $4.25^{+0.50}_{-0.61}\times10^{-1}$ & $1.08^{+0.10}_{-0.09}\times10^{4}$ & $18.09^{+0.09}_{-0.15}$ & $1.40^{+0.47}_{-0.26}$ & $1.28^{+0.37}_{-0.18}\times10^{4}$ & $1.10^{+0.36}_{-0.37}\times10^{3}$\\  
30-70 & $16.87^{+0.04}_{-0.04}$ & $3.49^{+0.32}_{-0.33}\times10^{-1}$ & $7.80^{+0.45}_{-0.48}\times10^{3}$ & $18.16^{+0.08}_{-0.14}$ & $1.26^{+0.42}_{-0.19}$ & $9.04^{+2.34}_{-1.10}\times10^{3}$ & $2.79^{+1.83}_{-1.78}\times10^{2}$\\  
70-100 & $16.93^{+0.03}_{-0.03}$ & $2.98^{+0.23}_{-0.24}\times10^{-1}$ & $3.50^{+0.18}_{-0.18}\times10^{3}$ & $18.12^{+0.10}_{-0.17}$ & $1.40^{+0.61}_{-0.30}$ & $4.44^{+1.67}_{-0.75}\times10^{3}$ & $55.83^{+58.62}_{-39.54}$\\  
100-150 & $17.01^{+0.03}_{-0.05}$ & $2.77^{+0.26}_{-0.26}\times10^{-1}$ & $3.21^{+0.18}_{-0.18}\times10^{3}$ & $18.20^{+0.08}_{-0.13}$ & $1.37^{+0.52}_{-0.26}$ & $3.89^{+1.35}_{-0.63}\times10^{3}$ & $56.33^{+49.89}_{-38.18}$\\  
150-200 & $16.92^{+0.03}_{-0.05}$ & $2.65^{+0.24}_{-0.20}\times10^{-1}$ & $1.99^{+0.13}_{-0.13}\times10^{3}$ & $17.81^{+0.26}_{-0.56}$ & $2.75^{+3.33}_{-1.15}$ & $3.90^{+8.08}_{-1.68}\times10^{3}$ & $15.11^{+20.23}_{-10.95}$\\  
200-300 & $16.96^{+0.02}_{-0.04}$ & $2.33^{+0.19}_{-0.17}\times10^{-1}$ & $1.62^{+0.13}_{-0.13}\times10^{3}$ & $16.74^{+0.71}_{-0.52}$ & $11.02^{+6.15}_{-5.87}$ & $3.15^{+8.85}_{-2.44}\times10^{4}$ & $9.67^{+13.79}_{-7.04}$\\  
300-500 & $17.00^{+0.02}_{-0.02}$ & $2.09^{+0.19}_{-0.18}\times10^{-1}$ & $1.52^{+0.14}_{-0.13}\times10^{3}$ & $16.45^{+0.77}_{-0.34}$ & $14.29^{+4.90}_{-7.65}$ & $5.12^{+11.94}_{-4.30}\times10^{4}$ & $21.78^{+21.31}_{-14.96}$\\  
500-1000 & $16.99^{+0.04}_{-0.15}$ & $2.15^{+0.64}_{-0.47}\times10^{-1}$ & $3.63^{+0.73}_{-0.76}\times10^{2}$ & $16.45^{+0.66}_{-0.34}$ & $15.75^{+8.09}_{-7.46}$ & $2.28^{+21.32}_{-1.99}\times10^{4}$ & $19.69^{+13.15}_{-12.91}$\\  
\hline
\end{tabular}
\end{table*}

\subsection{Summary of the observation features}
Here we summarize the intriguing features from observation: 
(a) the overall light curve resembles a classic ``single pulse", which is wider and peaks later in lower energies in a self-similar way; and (b) there is no underlying elementary single slow pulse and the apparent broad {\yisx profile} is composed with many fast pulses. 

\begin{figure}
    \centering
    \includegraphics[width=8cm]{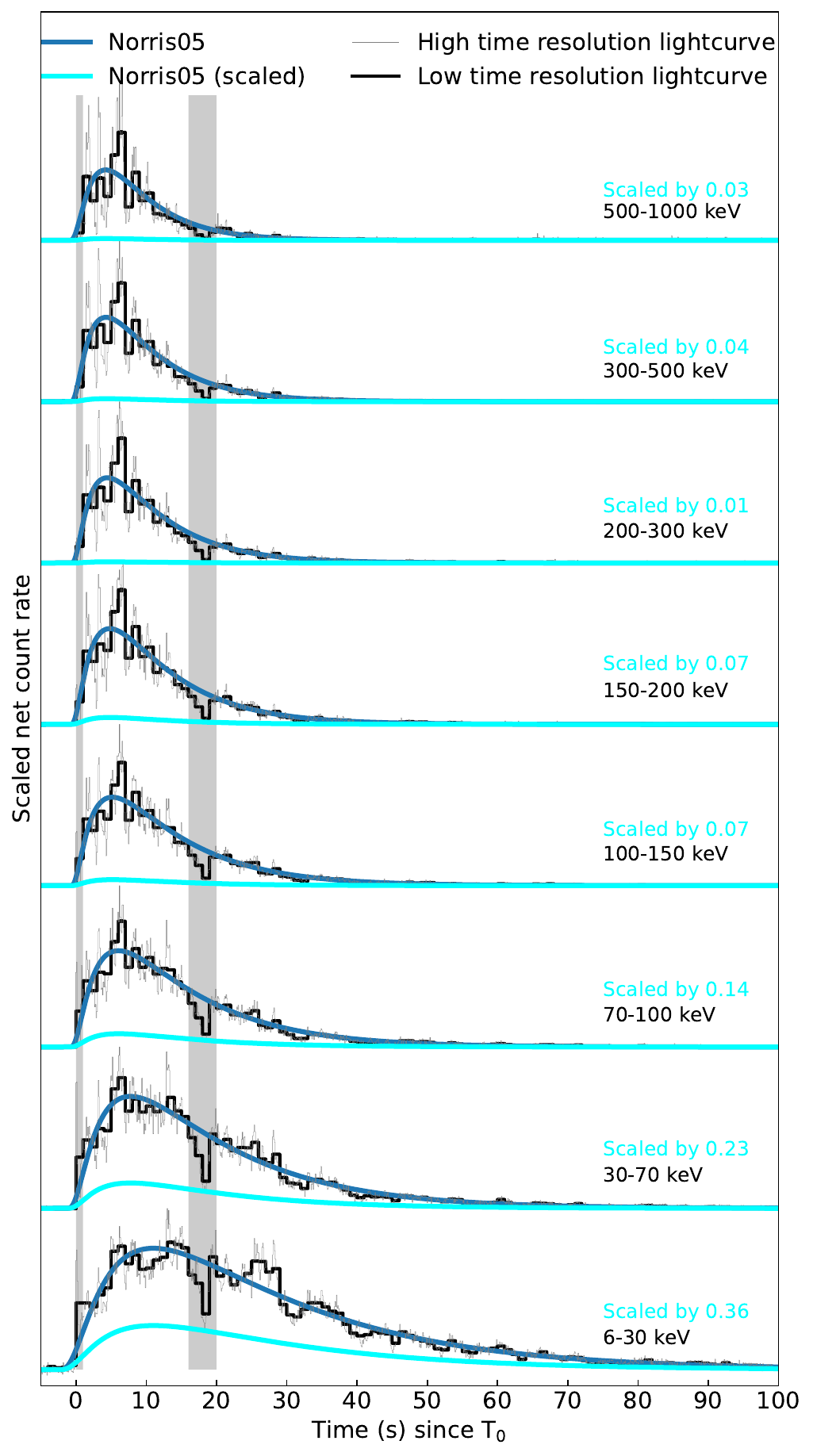}
    \caption{Maximum allowed underlying elementary broad pulses in each energy bands}
    \label{fig:max-allowed}
\end{figure}

In order to better demonstrate that the observed profile-energy dependence cannot be attributed to an underlying broad pulse, we perform the following analysis: we downscale the best-fit FRED-profiles in each energy bands, so that there are no negative residuals in the high time-resolution light curves. The rescaled broad FRED are the maximum allowed underlying broad pulses in each energy bands, we plot it in the Figure \ref{fig:max-allowed}. The found rescaling factor is the fraction of fluence attributed from this broad pulse. We can see that the rescaling factor in most energy bands are found to be neglegible, suggesting that the broad pulse-energy dependence cannot be attributed to an underlying broad pulse.

In other words, {\yisxapj the individual pulses work in concert with each other to make the overall profile-energy dependence.} 

\section{Physical implications}
The above features found in this burst indicate that individual fast pulses are correlated in time and energy. The implication of that is, the emission sites of individual pulses should be connected casually. It provides significant insights into the prompt emission mechanism and the composition of the jet. 
\subsection{The ICMART model}
\begin{figure}
    \centering
    \includegraphics[width=8 cm]{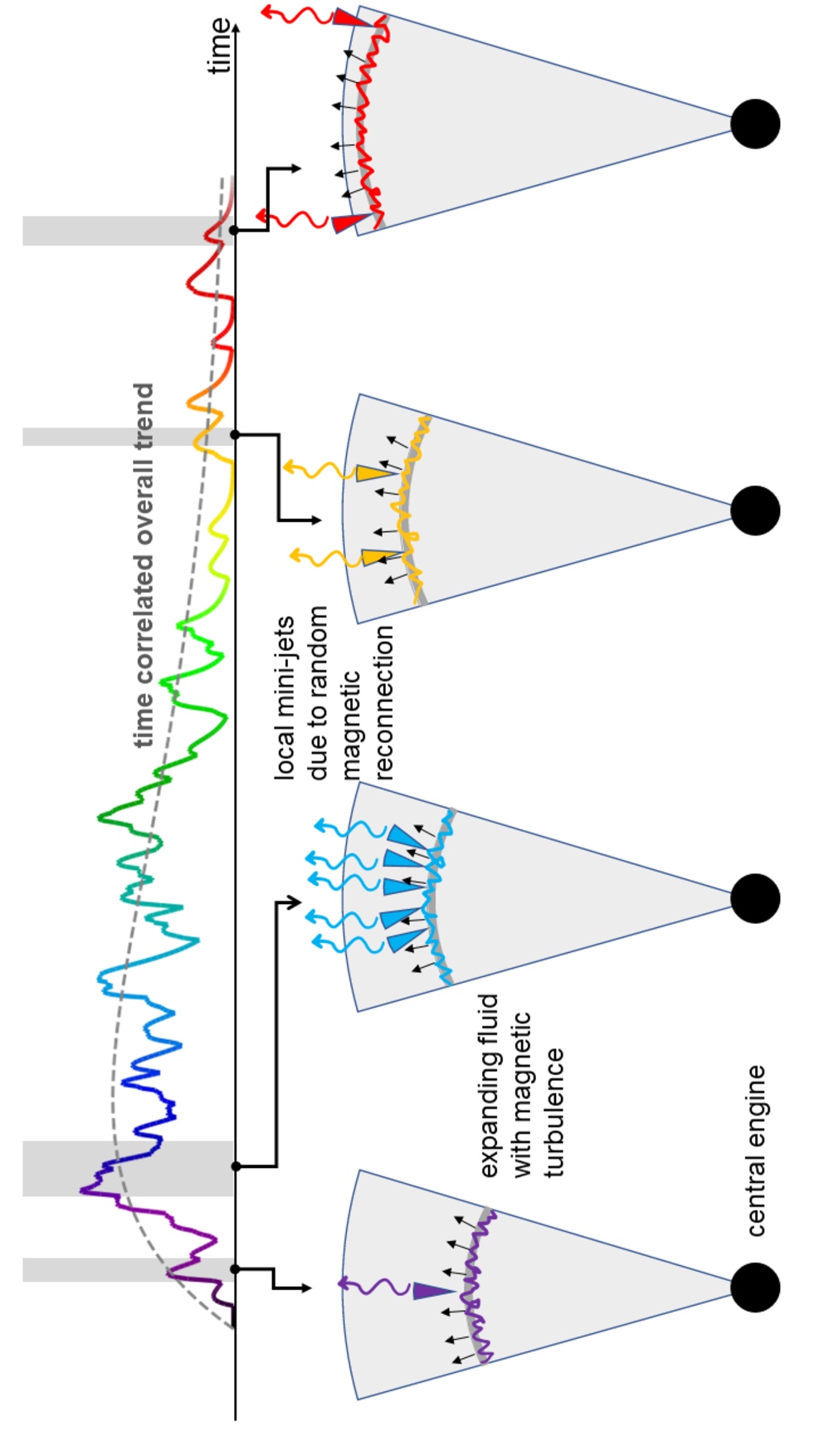}
    \caption{\textbf{An illustration of the ICMART model}. The overall trend of the light curve corresponds to the growth and decline of the number of magnetic reconnection events within a single magnetized fluid. The turbulence in the fluid triggers random local reconnection events which emits through mini-jets. Each mini-jet emission corresponds to a fast pulse observed in the light curve. Because the emission originates from a single fluid that travels in space, a global shape-energy dependence is naturally expected. Deep dips are allowed in occasions when no bright mini-jets beam towards Earth at a particular epoch.}
    \label{ill:ICMART}
\end{figure}

{\yisx In the ICMART model \citep{Zhang-Yan11,Zhang-Zhang14,Shao-Gao22}, individual fast pulses are emitted from mini-jets launched by local magnetic reconnections. All mini-jets are in the same expanding magnetically-dominated fluid (see illustration in Figure \ref{ill:ICMART})}. The rise and fall of the FRED-like profile is mostly dictated by the rate of growth of the number of reconnection events \citep{2024arXiv241116174Y}. During the evolution of the broad pulse, the magnetic field strength decays as the whole fluid expands \citep{Uhm-Zhang14b}. It leads to rolling down of the characteristic synchrotron emission frequency with time \citep{Uhm-Zhang16,Uhm-Zhang18}. The soften of the spectra over time leads to the observed profile-energy dependence \footnote{See \cite{Sunhui23} for the spectrum time evolution analysis of this burst.}. With certain spectrum evolution, a self-similar profile may be reproduced \citep{2024arXiv241116174Y}.  Within this picture, the observed prompt emission originates from a single episode of central engine activity, which is consistent with the merger origin of GRB 230307A \citep{Gillanders23, Sunhui23,yangtroja24}. 

\begin{figure}
    \centering
    \includegraphics[width=\linewidth]{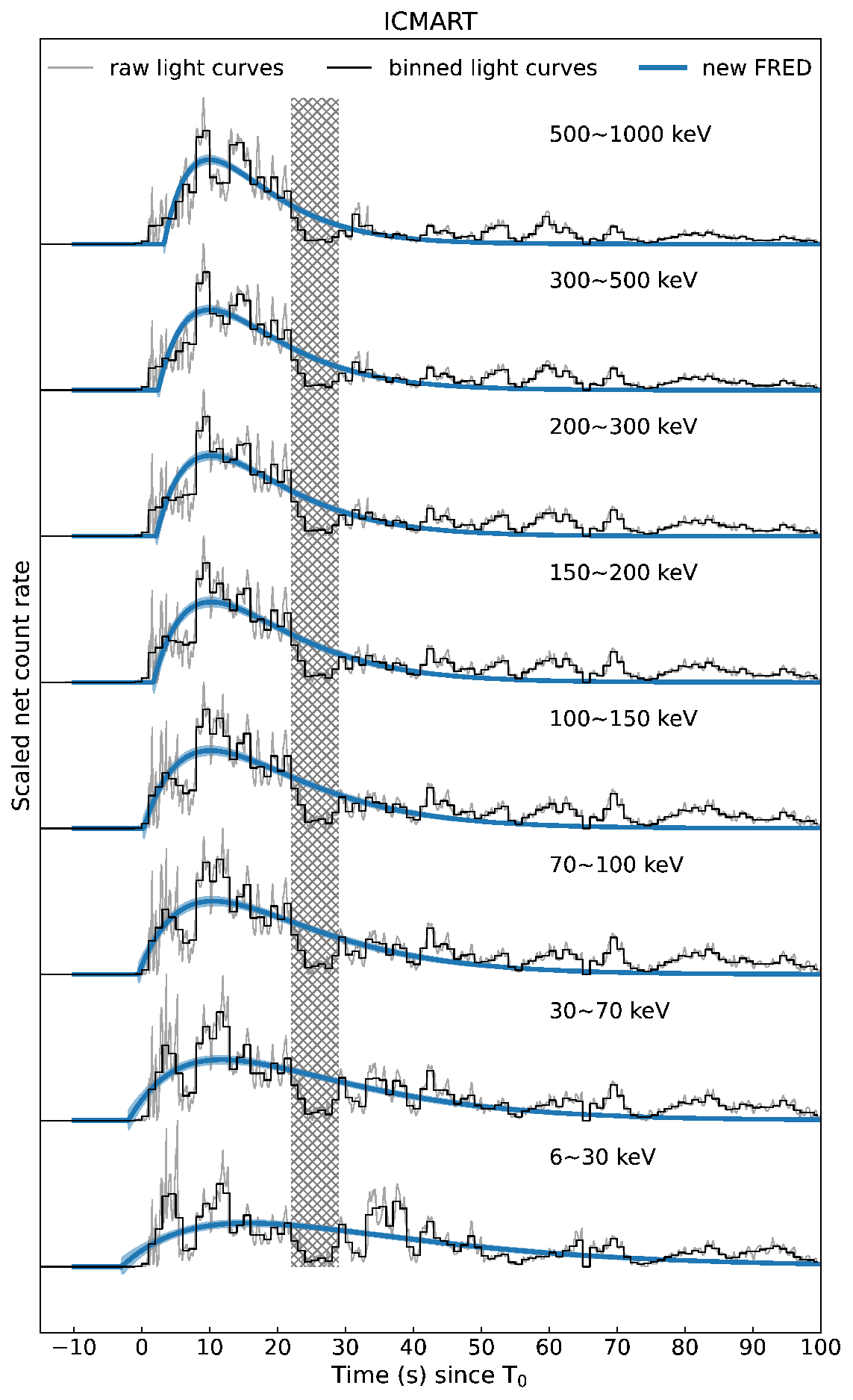}
    \caption{ {\bf The simulated multi-band light curves within the framework of ICMART model.} Both aligned short pulses and a global shape-energy dependence can be reproduced. The shaded area indicates the location of a reproduced light curve dip due to the lack of mini-jet emission at that time.}
    \label{fig:12}
\end{figure}

{\yisx In order to better illustrate the idea, we conduct ICMART simulations and present the results. The general settings of the simulation are as follows:} We assume a pair of magnetized shells collide and merge as one fluid after a collision at $\sim 10^{15}$\,cm. The mass of the fluid is assumed to be $10^{31}$\,g, the initial Lorentz factor is $30$ the initial magnetization factor is $\sigma = 50$ (i.e. the comoving-frame magnetic energy is $50$ times of the rest-mass energy of the fluid).
The number of reconnection events increases by a factor of $5$ during the emission period and the total number of the reconnection events within the $1/\Gamma$ cone is assumed to be $200$. The relatively small number of reconnection events was chosen to ensure the occurrences of dips in most of the realizations. Each reconnection event has a size of $\sim 4\times10^{13}$\,cm, a random latitude within $(0, 1/3)$ and a random orientation within $(0, \pi / 2)$. {\yisx For each reconnection, the electrons are assumed to be accelerated to power law distribution with index -2.8. The electrons cool down and radiate via synchrotron radiation. We calculate the cooling and radiation process according to \cite{Uhm-Zhang14b}, and superpose the resulting synchrotron spectrum of each mini-jet to obtain the observed spectrum.}
In Figure\,\ref{fig:12}, we {\yisx show the light curves reproduced by the simulation} that is consistent with the {\yisx key }observation {\yisx features}: 1) the light curve can be decomposed of many short pulses related to mini-jet emission; 2) the overall shape of the light curve resembles a single pulse, but can possess a very deep dip; 3) the overall single slow pulse has a significant shape evolution across energy bands. This simulation demonstrates that the ICMART model can well reproduce the observed behaviour of GRB 230307A. Detailed simulations will be presented in a separate paper. 

\subsection{Standard internal shock model}
\begin{figure}
    \centering
    \includegraphics[width=12 cm, angle=90]{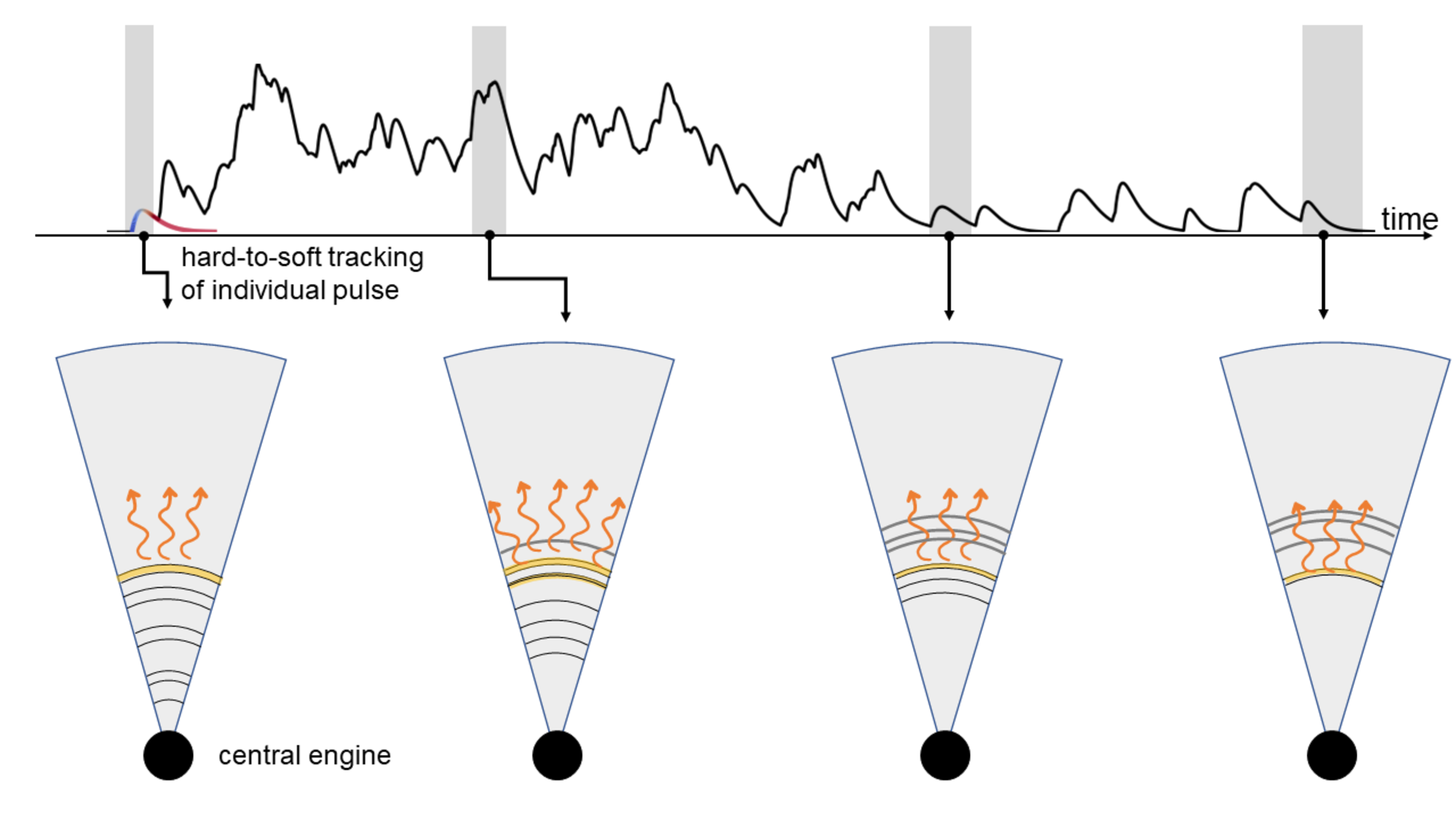}
    \caption{\textbf{An illustration of the standard internal shock picture.} Lower panels arranged from left to right depict the progression from earlier to later epochs. Each shock between a pair of shells (highlighted in lower panels) corresponds to a fast pulse in the light curve in the above panel (in corresponding shade regions). All collisions happen at similar distances from the central engine. Individual pulses may exhibit a spectral trend from hard to soft (illustrated with colour gradient). However, the overall light curve will not display any shape-energy dependence.}
    \label{ill:IS-A}
\end{figure}

The standard internal shock model, on the other hand, attributes short-time pulses as a consequence of collisions of many pairs of shells ({\revision see Figure \ref{ill:IS-A} for an illustration}). Within such a scenario, the overall trend is {\yisx determined by} the history of the central engine activity. Light curves in different bands should share this same history and should not display a global energy-dependent behaviour {\yisx as observed}.

\begin{figure}
    \centering
    \includegraphics[width=\linewidth]{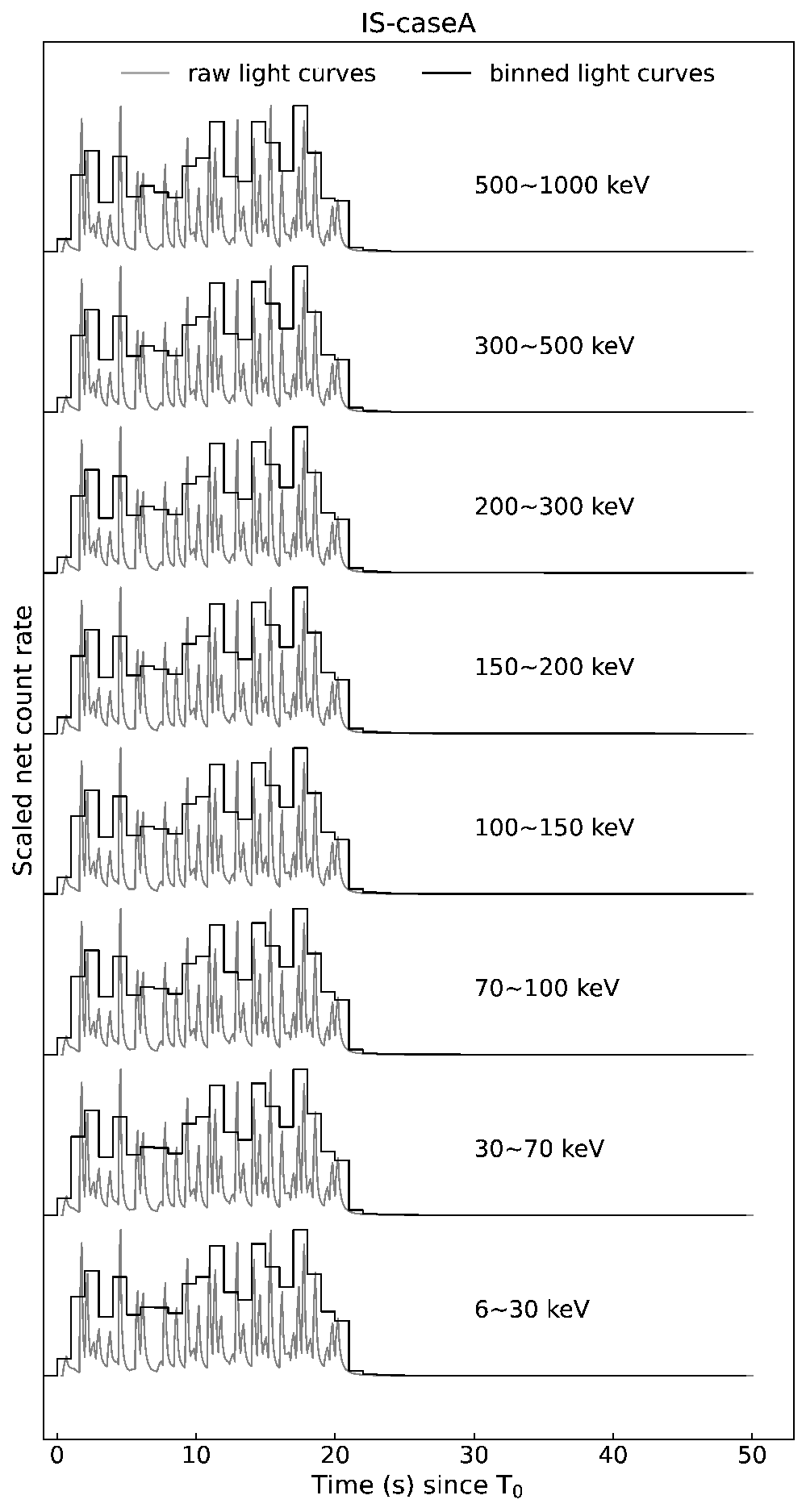}
    \caption{ {\bf The simulated multi-band light curves within the framework of the standard internal shock model.} It shows a random behaviour and the lack of global shape-energy dependence.}
    \label{fig:13}
\end{figure}

To better demonstrate this, we conducted simulations within the framework of the standard IS model with the following settings: We simulated the ejection of 50 shells from the central engine, with their initial thicknesses, Lorentz factors and masses sampled from random distributions in the logarithmic space. The Lorentz factors vary from 100 to 1000, the mass ranges from $10^{28}$ to $10^{29}$, and the initial thickness ranges from $10^{10}$ to $2 \times 10^{10}$, all in cgs units. The central engine's ejection times are randomly selected from a linear distribution, with values ranging from $0$ to $20$ seconds in the rest frame of the GRB central engine. Upon each collision, two shells merge, and the merged shell parameters ($m$, $\gamma$, $l_r$) are adopted as the new values for the remaining shell. To ensure tracking future collisions, the ``effective" ejection time of this new shell is redefined as $t_{\rm{ej},m}=R/c\beta_m$, where $\beta_m=(1-1/\gamma_m^2)^{1/2}$ and $\gamma_m$ is the Lorentz factor of the merged shell. The simulation then restarts with one less shell. This process is repeated for each collision, allowing the code to monitor all collision/merging events for any arbitrarily designed central engine activity \citep{Kobayashi97,Maxham-Zhang09}. {\yisx The spectra for each pulse are simulated empirically (see details in a separate paper \citep{2024arXiv241017189M})}. The results of the simulation is presented in Figure\,\ref{fig:13}, where there is no shape-energy dependence of the overall light curve as we expected. In the IS model, the energy-dependent lag and width generally apply to each pulse. 
Even if there is a stretching factor in low energies to make pulses broader, the stretching factor scales with the duration of short pulses. There is no way to reproduce a large stretching factor in the scale of the overall profile by superposing these short pulses.
\subsection{Modified internal shock model with another internal shock at large radius}
\begin{figure}
\centering
\includegraphics[width=8cm]{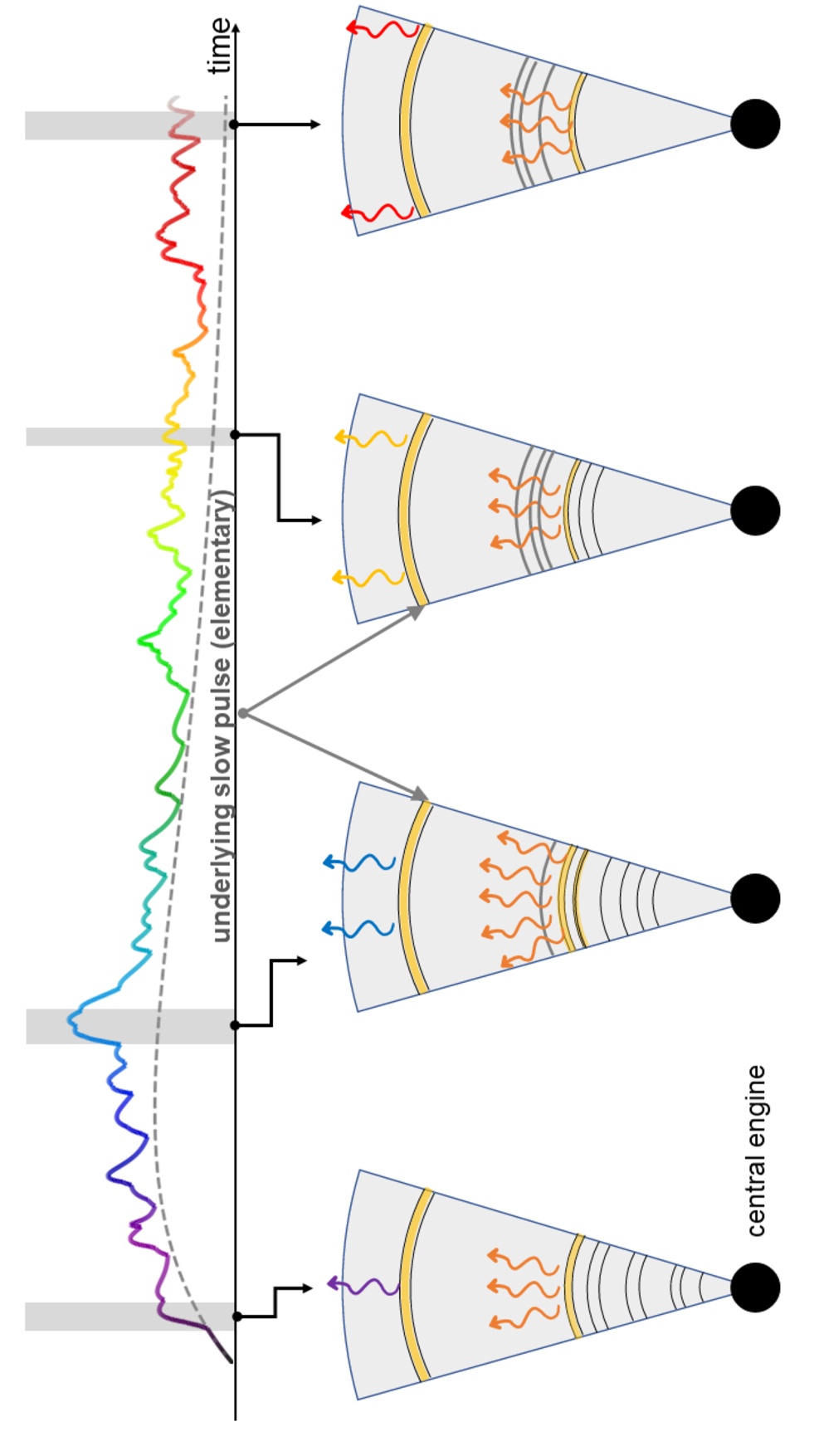}
\caption{\revision\textbf{An illustration of a modified internal shock model that invokes an additional collision at a large radius from the central engine.} In this case, the detected emission is a superposition of emission from fast pulses and the broad pulse. The overall light curve may indeed show shape-energy dependence, or equivalently an overall hard-to-soft trend (illustrated with colour gradient). However, the existence of an underlying slow pulse component forbids any dip features in the light curve.}
\label{ill:IS-B}
\end{figure}

A modified internal shock model (see Figure \ref{ill:IS-B}) may invoke another internal shock at a much larger emission radius \citep{1998MNRAS.296..275D,2003MNRAS.342..587D,2014A&A...568A..45B} defined by $R_2 \sim \Gamma^2 c \Delta t \sim (1.2\times 10^{16} \ {\rm cm}) (\Gamma/100)^2 (\Delta t / 40 \ {\rm s})$, where $\Delta t \sim 40$ s is the timescale of the broad pulse. However, within such a picture, the observed emission should be the superposition between the fast and slow components, which requires that fast pulses only grow on top of a {\yisx dominating} the slow component. {\yisx It is not supported by the existence of the very deep, achromatic dip at around 18\,s.} 

\begin{figure}
    \centering
    \includegraphics[width=\linewidth]{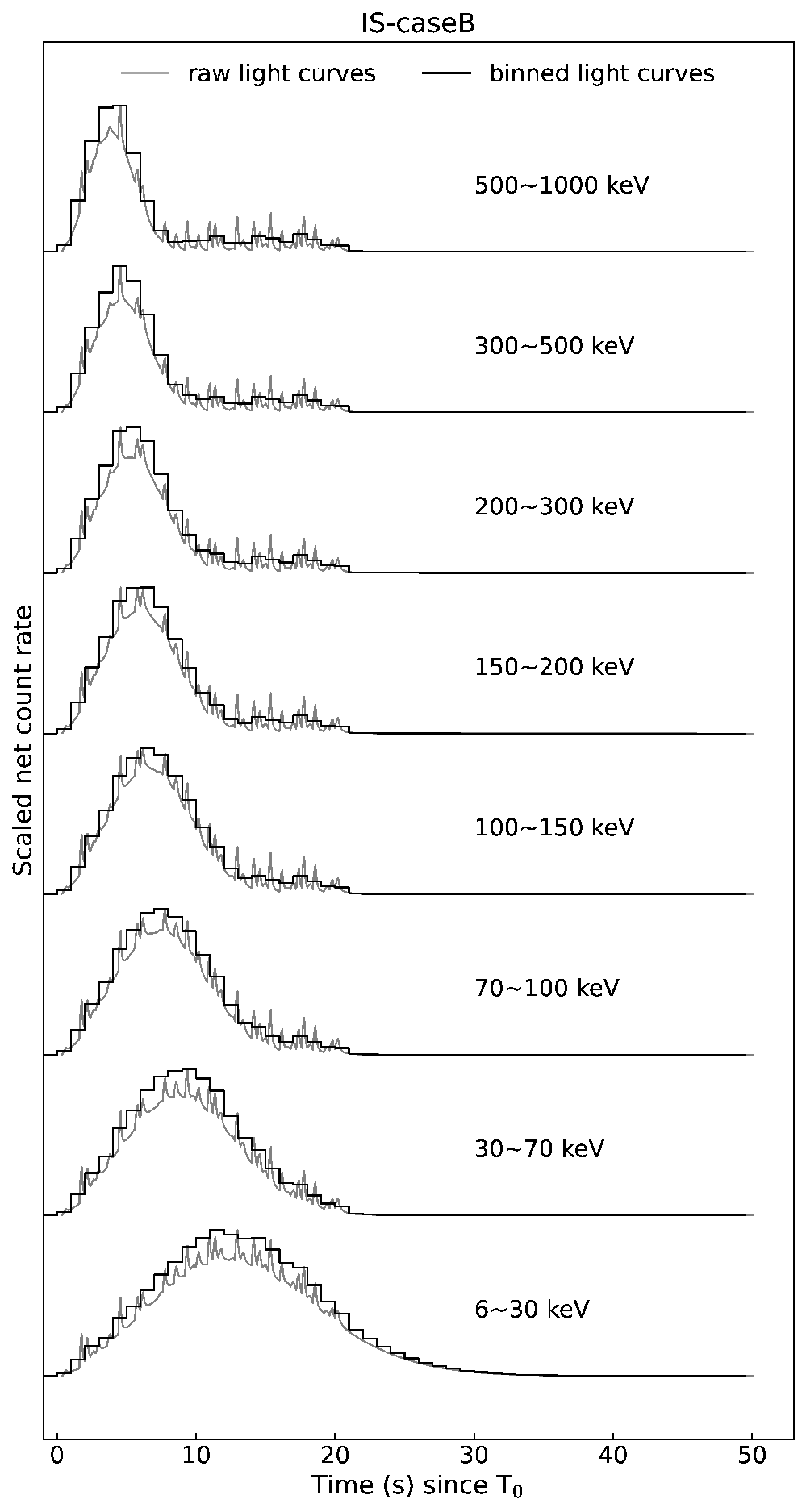}
    \caption{{\bf The simulated multi-band light curves within the framework of the modified internal shock model that invokes superposition a large-radius internal shock and many small-radii internal shocks.} The overall light curve is dominated by the emission from the large-radius internal shock, with no hope to produce dips in the energy-dependent light curves.}
    \label{fig:14}
\end{figure}

We also conduct a simulation under this variant IS model: a separate shock at a larger distance ($\sim 4 \times 10^{16} \rm cm$) is generated to make a slow pulse component that overlaps with those fast pulses. {\yisx We consider synchrotron radiation process in IS model to determine the spectrum evolution of the long pulse, using the analytical solution proposed by \cite{2003MNRAS.342..587D}, involving a relativistic wind where a slower shell ($\Gamma = 100$) decelerates a faster portion of the flow ($\Gamma_0 = 400$) characterized by a constant rest-frame mass flux.} The overall light curve is dominated by the slow pulse and shows the shape-energy dependence (as illustrated in Figure \,\ref{ill:IS-B}). However, in this case, there should be an {\yisx dominating} underlying slow pulse emission component, which is in direct contradiction with the observations presented in this work. The corresponding simulation results are presented in Figure\,\ref{fig:14}, which does not reproduce the observations, especially the deep dip. 

{\yisx In summary, any other attempt trying to model the prompt emission of this bursts within such frameworks should manage to address all the aforementioned challenges.}
\section{Summary}
In this paper, we demonstrate that the light curve of GRB 230307A can be decomposed into several individual pulses, while the overall behavior of these individual pulses (slow varying trend) resembles a self-similar broad single pulse. While the rapid variability shows alignment in all energy bands, the broad pulse has a well-defined energy-dependence manner. In order to reach the above observational conclusion solidly, we conduct the following analysis: 
\begin{itemize}
    \item We fit parameterized FRED profiles to the overall light curves in different energy bands, and found that the width and peak time of the profile decrease with the energy in a power law. This dependence was found to saturate at $\sim200$ keV;
    \item We found that the smoothed overall shape of the light curve in each band is essentially a time-stretched copy of that in the other bands; 
    \item After subtraction of the over all trend of the light curve, we show that the residual variations superimposed on the best-fit "self-similar" temporal profile do not undergo the same temporal stretching, but are instead fixed in time regardless of the band.
    \item We found that some prominent fast components can be fitted with individual pulses, such as the pulse at around 3.5\,s, and the dip at around 18\,s, without the need of an underlying slow components; 
    \item The deep dip at around 18\,s significantly constrained any possible elementary slow components, to which the observed overall profile-energy dependence can be attributed. 
\end{itemize}
The above mentioned features indicate that the prompt emission of this burst is from many individual emitting units that are casually linked in a emission site at a large distance from the central engine. We elaborated and demonstrated with simulations that, this scenario is consistent with the picture of many mini-jets due to local magnetic reconnection events in a large emission zone far from the GRB central engine (as envisaged in the ICMART model) but raises a great challenge to the internal shock-like models that attribute all variability components to collisions among different shells. 

\bibliography{main}{}
\section{Acknowledgement} 
S.-X.Y. thanks the insightful discussion with Prof. Kinwah Wu on the origin of the self-similarity of the multi-wavelength light curves of the burst. S.-X.Y. also discussed with Dr. X. L. Wang and Dr. X. Y. Song on the potential physical implication of these interesting behaviour in data. This work is supported by the National Key R\&D Program of China (2021YFA0718500). S.-X.Y. acknowledges support from the Chinese Academy of Sciences (grant Nos. E329A3M1 and E3545KU2). S.-L.X. acknowledges the support by the National Natural Science Foundation of China (Grant No. 12273042). The GECAM (Huairou-1) mission is supported by the Strategic Priority Research Program on Space Science (Grant No. XDA15360000) of Chinese Academy of Sciences. 


\end{document}